\documentclass[twocolumn,twocolumnappendix]{aastex63}
\usepackage{mathtools}
\usepackage[flushleft]{threeparttable}
\usepackage{graphics,graphicx,subfigure}
\usepackage{epstopdf}
\usepackage{epsfig}
\usepackage{helvet}
\usepackage[utf8]{inputenc}
\usepackage[T1]{fontenc}
\usepackage{longtable}
\usepackage{soul}
\usepackage{amsmath, amssymb, gensymb} 
\usepackage{natbib}
\usepackage{microtype}
\usepackage{multirow}
\usepackage{float}
\usepackage{morefloats}
\usepackage{outlines}
\usepackage{color}
\usepackage{appendix}
\usepackage{xspace}
\usepackage{CJK}
\defcitealias{Kral2016}{K16}

\def\mjybm {mJy\,beam$^{-1}$\xspace}
\def\ujybm {$\mu$Jy\,beam$^{-1}$\xspace}

\def\etal {\textit{et al.}\xspace}

\def\betapic {$\beta$~Pic\xspace}

\def\minoraxis {--59}
\def\majoraxis {31}

\def\resolution {$0\farcs97$\xspace}
\def\pixel{0$\farcs$14\xspace}

\def\boxmajorpix {114}
\def\boxminorpix {12}
\def\boxmajorarcsec {16$\farcs$0}
\def\boxminorarcsec {1$\farcs$7}
\def\boxmajorau {310}
\def\boxminorau {32.7}

\def\boxdividedmajorpix {130}
\def\boxdividedminorpix {12}
\def\boxdividedmajorarcsec {18$\farcs$2}
\def\boxdividedminorarcsec {1$\farcs$7}
\def\boxdividedmajorau {354}
\def\boxdividedminorau {32.7}

\def\beamsInBox {30}

\def\rmsI {28.1}
\def\rmsQ {13.0}
\def\rmsU {12.6}
\def\rmsQU {12.8}
\def\rmsP {18.2}

\def\Pfraclimits {1.6}
\def\Pfraclimitsfolded {1.1}

\def\peakI {2.59}
\def\fluxI {47}

\def\Iboxavg {1368}
\def\Qboxavg {--4.0}
\def\Uboxavg {--6.9}
\def\Pboxavg {6.9} 
\def\Qprimeboxavg {8.0}
\def\Uprimeboxavg {--1.5}
\def\rmsQUboxavg {2.6}
\def\rmsPboxavg {3.6} 
\def\SNRQboxavg {1.5\xspace}
\def\SNRUboxavg {2.7\xspace}
\def\SNRPboxavg {2.7\xspace} 
\def\SNRQprimeboxavg {3.1\xspace}
\def\SNRUprimeboxavg {0.1\xspace}
\def\Pfracboxavg {0.0051\,$\pm$\,0.0019\xspace}
\def\Pfracpctboxavg {0.51\,$\pm$\,0.19}
\def\PAboxavg {--59.9\,$\pm$\,10.6}
\def\PAprimeboxavg {--0.5\,$\pm$\,10.6}

\def\Pfracouter {1.1}
\def\SNRPouter {2.1\xspace}
\def\SNRQpouter {2.3\xspace}

\def\PfracSWthird {1.1}
\def\SNRPSWthird {2.4\xspace}
\def\SNRQpSWthird {2.8\xspace}

\received{21 Jan 2022}
\revised{10 Mar 2022}
\accepted{21 Mar 2022}
\submitjournal{ApJ}

\shorttitle{$\beta$ Pic Dust Polarization}
\shortauthors{Hull \etal}
\graphicspath{{figures/}}

\begin{document}
\begin{CJK*}{UTF8}{gbsn}

\title{Polarization from Aligned Dust Grains in the $\beta$ Pic Debris Disk}

\correspondingauthor{Charles L. H. Hull, Haifeng Yang}
\email{chat.hull@nao.ac.jp, hfyang@pku.edu.cn}

\author[0000-0002-8975-7573]{Charles L. H. Hull}
\altaffiliation{NAOJ Fellow}
\affiliation{National Astronomical Observatory of Japan,
      Alonso de C\'{o}rdova 3788,
      Office 61B,
      7630422,
      Vitacura,
      Santiago,
      Chile}
\affiliation{Joint ALMA Observatory,
      Alonso de C\'{o}rdova 3107,
      Vitacura,
      Santiago,
      Chile}

\author[0000-0002-8537-6669]{Haifeng Yang (杨海峰)}
\affil{Kavli Institute for Astronomy and Astrophysics, Peking University, Yi He Yuan Lu 5, Haidian Qu, Beijing 100871, China}
\affil{Institute for Advanced Study, Tsinghua University, Beijing, 100084, China}

\author[0000-0002-3583-780X]{Paulo C. Cort\'es}
\affil{National Radio Astronomy Observatory, 520 Edgemont Road, Charlottesville, VA 22903, USA}
\affiliation{Joint ALMA Observatory,
      Alonso de C\'{o}rdova 3107,
      Vitacura,
      Santiago,
      Chile}
      
\author{William R. F. Dent}
\affil{European Southern Observatory, Alonso de C\'ordova 3107, Vitacura, Santiago, Chile}
\affiliation{Joint ALMA Observatory,
      Alonso de C\'{o}rdova 3107,
      Vitacura,
      Santiago,
      Chile}

\author[0000-0001-6527-4684]{Quentin Kral}
\affil{LESIA, Observatoire de Paris, Universit\'e PSL, CNRS, Sorbonne Universit\'e, Universit\'e Paris Diderot, Sorbonne Paris Cit\'e, 5 place Jules Janssen, 92195, Meudon, France}

\author[0000-0002-7402-6487]{Zhi-Yun Li}
\affiliation{Department of Astronomy, University of Virginia, Charlottesville, VA 22903, USA}

\author[0000-0002-5714-799X]{Valentin J. M. Le Gouellec}
\affil{SOFIA Science Center, Universities Space Research Association, NASA Ames Research Center, Moffett Field, California 94035, USA}
\affil{Universit\'e Paris-Saclay, CNRS, CEA, Astrophysique, Instrumentation et Mod\'elisation de Paris-Saclay, F-91191, Gif-sur-Yvette, France}
\affil{European Southern Observatory, Alonso de C\'ordova 3107, Vitacura, Santiago, Chile}

\author[0000-0002-4803-6200]{A. Meredith Hughes}
\affil{Department of Astronomy, Van Vleck Observatory, Wesleyan University, 96 Foss Hill Drive, Middletown, CT 06459, USA}

\author[0000-0001-9325-2511]{Julien Milli}
\affil{Universit\'e Grenoble Alpes, IPAG, F-38000 Grenoble, France}

\author[0000-0003-1534-5186]{Richard Teague}
\affil{Center for Astrophysics $\vert$ Harvard \& Smithsonian, 60 Garden Street, Cambridge, MA 02138, USA}

\author[0000-0001-9064-5598]{Mark C. Wyatt}
\affil{Institute of Astronomy, University of Cambridge, Madingley Road, Cambridge CB3 0HA, UK}


\begin{abstract}

\noindent
We present 870\,$\micron$ ALMA polarization observations of thermal dust emission from the iconic, edge-on debris disk \betapic.  While the spatially resolved map does not exhibit detectable polarized dust emission, we detect polarization at the $\sim$\,3\,$\sigma$ level when averaging the emission across the entire disk. The corresponding polarization fraction is $P_\textrm{frac}$\,=\,\Pfracpctboxavg\%. The polarization position angle $\chi$ is aligned with the minor axis of the disk, as expected from models of dust grains aligned via radiative alignment torques (RAT) with respect to a toroidal magnetic field ($B$-RAT) or with respect to the anisotropy in the radiation field ($k$-RAT).  
When averaging the polarized emission across the outer versus inner thirds of the disk, we find that the polarization arises primarily from the SW third.
We perform synthetic observations assuming grain alignment via both $k$-RAT and $B$-RAT. Both models produce polarization fractions close to our observed value when the emission is averaged across the entire disk. When we average the models in the inner versus outer thirds of the disk, we find that $k$-RAT is the likely mechanism producing the polarized emission in \betapic.
A comparison of timescales relevant to grain alignment also yields the same conclusion.
For dust grains with realistic aspect ratios (i.e., $s > 1.1$), our models imply low grain-alignment efficiencies.
\end{abstract}

\keywords{ 
\textit{(Unified Astronomy Thesaurus concepts)} 
A stars (5);
Debris disks (363);
Dust continuum emission (412);
Interferometry (808);
Interplanetary dust (821);
Interplanetary magnetic fields (824);
Polarimetry (1278);
Submillimeter astronomy (1647);
Radiative transfer simulations (1967);
Theoretical models (2107)
}


\section{Introduction} \label{sec:intro}

One of the long-standing goals of disk enthusiasts has been to make a well resolved map of the magnetic field in a protoplanetary disk.  The detection of polarization in a disk from dust grains aligned with the magnetic field \citep{Lazarian2007} as a result of the Radiative Alignment Torque mechanism (or RAT; \citealt{LazarianHoang2007a}) would provide evidence that young protostellar disks are magnetized; this is a prerequisite for the operation of the magneto-rotational instability (MRI; \citealt{Balbus1991}) and magnetized disk winds \citep{Blandford1982}, both of which are thought to play a crucial role in disk evolution. While there have been a few detections in the mid-infrared of polarization from magnetically aligned dust grains in disks around Herbig Ae/Be stars \citep{DLi2016,DLi2018}, most of the ALMA observations of thermal dust polarization in disks, primarily at wavelengths of 1.3\,mm or 850\,$\micron$, can be interpreted as arising from scattering by dust grains \citep{Stephens2014, FernandezLopez2016, Kataoka2016b, Stephens2017b, Bacciotti2018, CFLee2018a, Girart2018, Hull2018a, SOhashi2018, Dent2019, Harrison2019, Vlemmings2019, SOhashi2020, Teague2021}, consistent with theoretical predictions \citep[e.g.,][]{Cho2007, Kataoka2015, Yang2016a}.  Observations of spectral-line polarization from the Goldreich-Kylafis effect \citep{Goldreich1981, Goldreich1982} offer an alternative way to probe the magnetic field in disks; however, searches for polarized spectral-line emission in bright, nearby Class II disks have thus far yielded either non-detections \citep{Stephens2020} or very low-level detections whose corresponding magnetic field morphologies were not easy to constrain \citep{Teague2021}.

The young, Class I and II protoplanetary disks toward which polarization from scattering is predominant are thought to be optically thick in the (sub)millimeter regime \citep{CarrascoGonzalez2016, CarrascoGonzalez2019,  Zhu2019}, which is conducive to producing a detectable level of scattering-induced polarization.  Consequently, the targets with the best potential for allowing us to minimize the contribution from scattering and detect polarization from magnetically aligned dust grains---thus allowing us to constrain the magnetic field in a solar system precursor---are debris disks, which are optically thin in continuum emission at millimeter wavelengths.  

Debris disks are tenuous, dust-dominated circumstellar disks analogous to the Solar System's Kuiper Belt and zodiacal light \citep{Hughes2018}.  The existence of debris disks was first inferred from observations by the IRAS satellite that showed infrared excesses around a number of stars including, among others, $\beta$ Pictoris (\betapic), $\alpha$ Lyrae (Vega), $\alpha$ Piscis Austrini (Fomalhaut), and $\epsilon$ Eridani \citep{Aumann1984, Aumann1985}.  \betapic\footnote{Throughout this paper we will use the name \betapic to refer both to the central star and to the debris disk surrounding it.}  was the first debris disk to be imaged at optical wavelengths \citep{Smith1984}.

\betapic is a main-sequence A-type (A6V) star with an effective temperature of 8052\,$K$ \citep{Gray2006}; a mass and radius of $1.797 \pm 0.035\,M_\odot$ and $1.497 \pm 0.25\,R_\odot$, respectively \citep{Zwintz2019}; and a bolometric luminosity or $8.7\,L_\odot$ \citep{Crifo1997}.  \betapic is located in the southern constellation of Pictor at a distance of $19.44 \pm 0.05$\,pc \citep[][using Hipparcos data]{vanLeeuwen2007}. The age of the \betapic moving group (and thus of \betapic itself) is calculated to be 
18.5$^{+2.0}_{-2.4}$\,Myr \citep{MiretRoig2020}.

\betapic hosts a large, bright, edge-on debris disk with a major axis extent of $\sim$\,3000\,au when observed at optical wavelengths \citep{Larwood2001}; the millimeter-wavelength observations that we present here (and others in the literature, e.g., \citealt{Wilner2011,Dent2014}), which trace larger dust grains, reveal a disk diameter of $\sim$\,300\,au.  The position angle of the major axis of the disk (measured east of north) is approximately \majoraxis$\degree$. 
The millimeter-wave dust emission has been resolved vertically (i.e., along the minor axis) by previous ALMA observations \citep[][see also Footnote \ref{footnote:scaleheight}]{Matra2019}.
Optical images of \betapic show a warped inner disk \citep[e.g.,][]{Mouillet1997,Heap2000,Apai2015}, which has been attributed to possible perturbations by planetary companions. And indeed, two giant planets have been directly observed orbiting the central star: \betapic~b \citep{Lagrange2010} and \betapic~c \citep{Lagrange2019, Lagrange2020, Nowak2020}. 

The \betapic debris disk is known to be gas rich. Emission has been detected from many atomic and molecular species including, e.g., CO and \ion{C}{1} at (sub)millimeter wavelengths \citep{Dent2014, Kral2016, Matra2017, Cataldi2018}; \ion{C}{2} and \ion{O}{1} in the far infrared \citep{Cataldi2014, Brandeker2016, Kral2016}; \ion{Na}{1}, \ion{Fe}{1}, and \ion{Ca}{2} in the optical \citep{Olofsson2001, Nilsson2012}; and CO, \ion{O}{1}, \ion{C}{1}, \ion{C}{2}, and \ion{C}{3} in the far ultraviolet \citep{Roberge2000, Roberge2006}.  
The total gas quantity in \betapic is lower than in protoplanetary disks; however, recent studies suggest that the gas may evolve viscously due to the MRI, which may be more active in debris disks than in protoplanetary disks and may operate in a different regime, i.e., one dominated by ambipolar diffusion \citep{KralLatter2016}. A better knowledge of the magnetic field in \betapic would help us better understand whether the MRI can indeed function and explain the system's gas distribution \citep{Kral2016}.

While there have been many optical and near-infrared observations of debris disks that probe polarization from Rayleigh or Mie scattering by small ($\sim$\,0.1--5$\,\micron$) dust grains, there is little evidence of polarization from aligned dust grains in debris disks.  To our knowledge, the only detection of such polarization is toward the debris disk BD~+31$\degree$643, whose polarized emission at optical wavelengths may be due to a combination of polarization from scattering and from dichroic extinction by dust grains aligned with a toroidal magnetic field in the disk \citep{Andersson1997}. In this work we use 870\,$\micron$ ALMA polarization observations of  thermal dust emission from \betapic to search for polarization from aligned dust grains. 

Below we describe our observations (\S\,\ref{sec:obs}) and our main results, which feature, most notably, a detection of polarized dust continuum emission when averaging over the entire disk of \betapic (\S\,\ref{sec:res}).  We continue with an exploration of dust-grain alignment models in an attempt to explain the low level of polarized emission that we see toward \betapic.  
We begin with an introduction to the debris disk model that we employ (\S\,\ref{sec:model}), first in a simple radiative transfer model that we use to constrain the expected intrinsic polarization fraction of emission from aligned dust grains given our observations (\S\,\ref{sec:rt}), and next in a more detailed synthetic observation, which fits our ALMA observations well (\S\,\ref{sec:synobs}).
We use the results from these models to constrain current models of dust-grain populations in debris disks (\S\,\ref{sec:constraints_dust}).
Next we explore our results in the context of grain-alignment theory (\S\,\ref{sec:theory}).     Finally, we discuss the implications of these findings and offer concluding thoughts (\S\,\ref{sec:conc}). 

\section{Observations and Imaging}
\label{sec:obs}

We present ALMA linear-polarization observations of \betapic taken at Band 7 (870\,$\micron$) with the 12\,m array. The data were taken in two sessions: first on 18 December 2019 under good weather conditions (precipitable water vapor [PWV] of 1.03--1.11\,mm and a phase rms of 19--22\,$\mu$m), with 42 antennas; and again on 19 December under improved conditions (PWV of 0.72--0.93\,mm and a phase rms of 18--59\,$\mu$m), with 43 antennas. During the observations the antennas were in the C-1 configuration, which has baseline lengths ranging from 14\,m to 312\,m. These baselines allow the recovery of emission up to angular scales as large as $\sim$\,7$\farcs$9 
and yield a synthesized beam (resolution element) in the combined dataset with dimensions of 1$\farcs$08\,$\times$\,0$\farcs$88 and a position angle of $-74.3\degree$ when imaged with a Briggs weighting parameter of \texttt{robust}\,=\,2.0 (i.e., natural weighting).  This average angular resolution of \resolution corresponds to a spatial resolution of approximately 19\,au at the distance to \betapic of 19.44\,pc.  

The following calibrators were included in the observations: J0522-3627 (polarization and pointing), J0538-4405 (bandpass and flux), and J0526-4830 (complex gain). The two sessions had sufficient parallactic angle coverage of the polarization calibrator (118$\degree$ and 116$\degree$ in the first and second sessions, respectively), which allowed us to perform polarization calibration (see below). The total on-source time was approximately 189\,min, yielding a thermal noise level (i.e., sensitivity) of \rmsQU\,\ujybm.

Our observations used the standard correlator configuration for ALMA Band 7 wide-band continuum (TDM) polarization observations, which includes 8.0\,GHz of bandwidth ranging in frequency from $\sim$\,337.5--341.5\,GHz and $\sim$\,347.5--351.5\,GHz, with an average frequency of 344.5\,GHz (870\,$\micron$).  Each 2\,GHz spectral window (with 1.875\,GHz of usable bandwidth) is divided into 64 channels with widths of 31.25\,MHz.

We obtained the raw data before the data were processed by the ALMA East Asian ALMA Regional Center (EA ARC), and thus we performed our own calibration of the data using version 5.6.1-8 of the Common Astronomy Software Applications (CASA; \citealt{McMullin2007}) by following recent versions of the standard ALMA pipeline and polarization calibration scripts. For a detailed description of the ALMA polarization calibration procedure, see \citet{Nagai2016}. The data were later reduced by the EA ARC staff, whose results matched our own.

In their combined study of Gaia and Hipparcos data, \citet{Snellen2018} derived the proper motion $\mu$ of \betapic: $\mu_\textrm{RA}$ = 4.94\,$\pm$\,0.02\,mas\,yr$^{-1}$ and  $\mu_\textrm{DEC}$ = 83.93\,$\pm$\,0.02\,mas\,yr$^{-1}$, very similar to the values derived from the Hipparcos-only data \citep{vanLeeuwen2007}.  The ICRS coordinates of \betapic measured from the second Gaia data release (DR2; \citealt{Gaia2018}) in epoch 2015.5 (i.e., its position on approximately 30 June 2015) were ($\alpha$, $\delta$) = (05:47:17.0960784\,$\pm$\,0.315\,mas, --51:03:58.132908\,$\pm$\,0.342\,mas) \citep{Snellen2018}.  Using the proper motions from \citeauthor{Snellen2018}, we can extrapolate from the Gaia DR2 position to the position of \betapic in our Band 7 data observed in mid-December 2019.  The resulting position matches the position of the central peak in our 2019 Band 7 image to within the $\sim$\,100\,mas positional uncertainty in our fits, which is due to the low resolution of our data and the extended nature of the emission toward \betapic. We thus choose to use the position of \betapic extrapolated from the proper motions as the position of the source (i.e., the position of the central star) at the time of our Band 7 observations.  That position in J2000 coordinates is ($\alpha$, $\delta$) = (05$^h$47$^m$17.098$^s$, --51$\degree$03$\arcmin$57$\farcs$758).  We use the CASA task \texttt{FIXVIS} to redefine the phase center of our observations, fixing it to the aforementioned position.  We then use the CASA task \texttt{FIXPLANETS} to set the coordinates of the phase center to be equal to that position; this latter step converts the coordinates from ICRS (the current default ALMA coordinate system) to J2000. 

Because the $\sim$\,15$\arcsec$ angular extent of \betapic nearly fills the $\sim$\,18$\arcsec$ field of view (also known as the ``primary beam''; the reported extent of the field of view represents the full-width at half-maximum, or FWHM) of the ALMA 12\,m antennas at Band 7, we observed \betapic in three separate pointings: one centered on the location of the central star \betapic and two located $\sim$\,3$\farcs$5 along the major axis in the NE and SW directions. When making our final images, we combine the data from all three pointings in a mosaic.  The 1\,$\sigma$ systematic uncertainty in polarization fraction for a single-pointing, on-axis linear polarization observations with ALMA is 0.03\% of Stokes $I$ (corresponding to a minimum detectable polarization fraction of 0.1\%).  The 1\,$\sigma$ systematic errors increase to $\sim$\,0.5\% near the FWHM of the primary beam in Band 7 observations; however, these errors are reduced by mosaicking \citep{Hull2020b, Cortes2021}.  Furthermore, the off-axis errors within the inner $\frac{1}{3}$ of the FWHM at Band 7 are at most 0.1\% \citep{Hull2020b}, and our small mosaic has been setup so that all of the emission from the \betapic disk falls within the inner $\frac{1}{3}$ of one of the three pointings; the systematic errors in polarization fraction are thus approximately 2\,$\times$ smaller than the statistical errors in our measurements (see Section \ref{sec:res} and Table \ref{table:data}).

\begin{figure}
    \centering
    \includegraphics[width=0.40\textwidth, clip, trim=0cm 0cm -0.9cm 0cm]{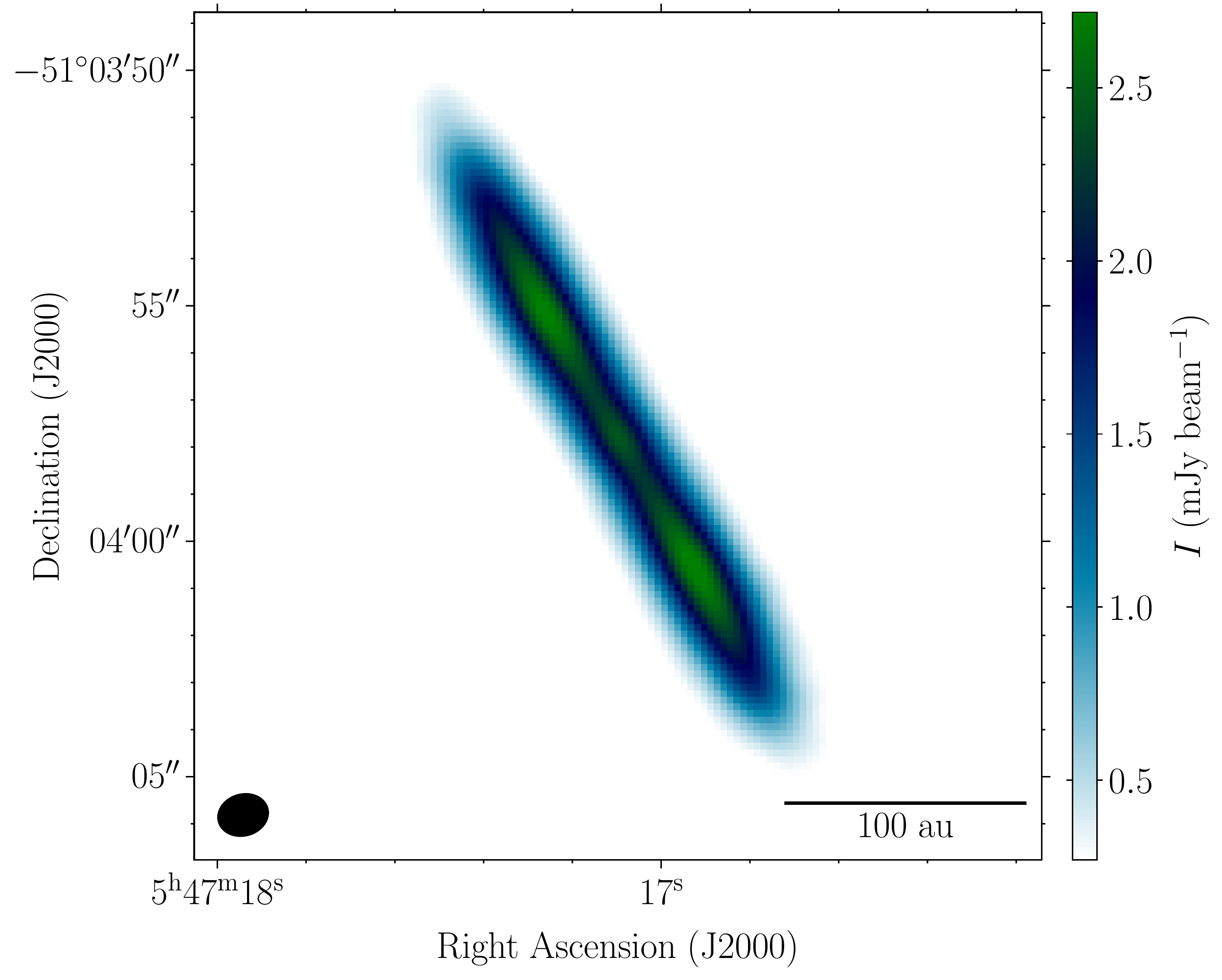}
    \includegraphics[width=0.40\textwidth]{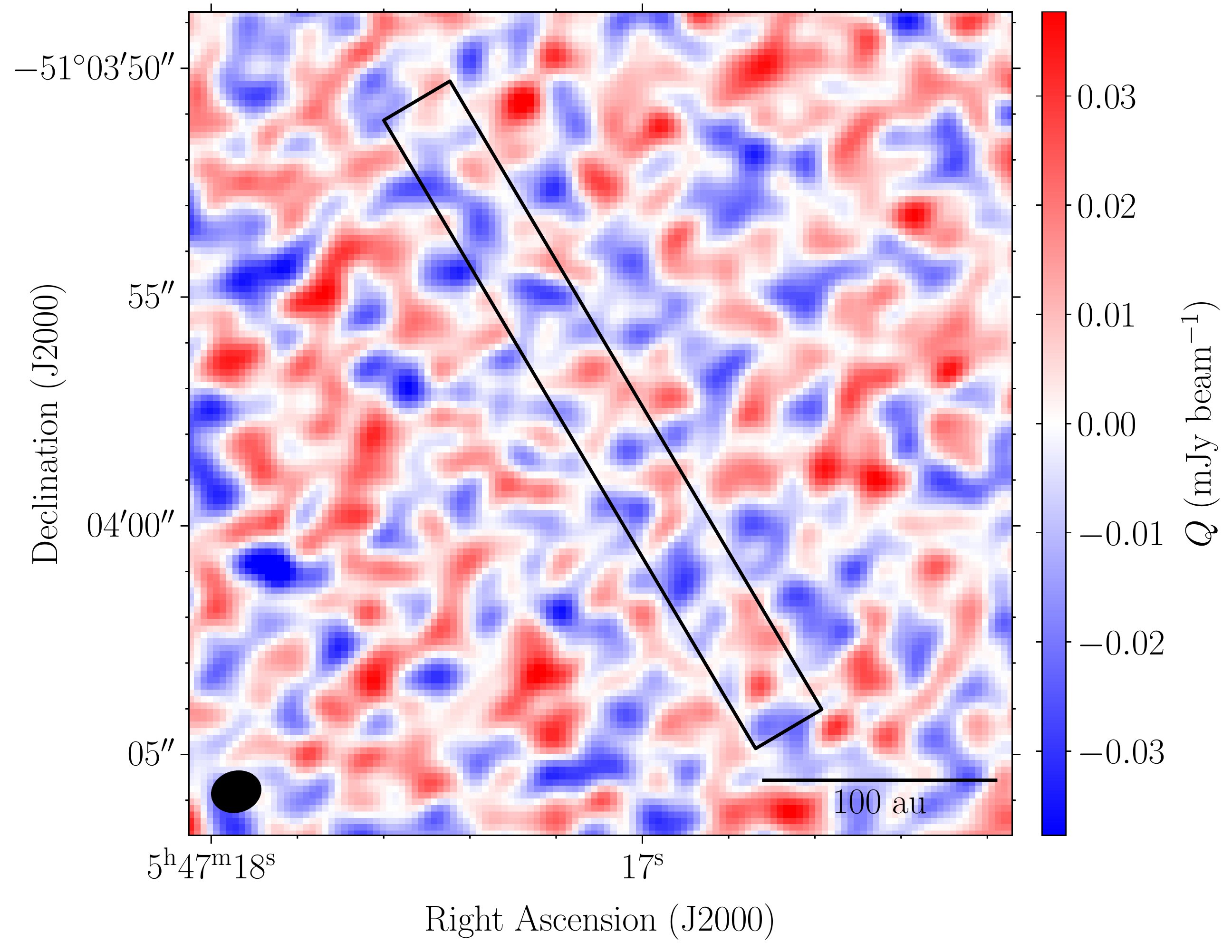}
    \includegraphics[width=0.40\textwidth]{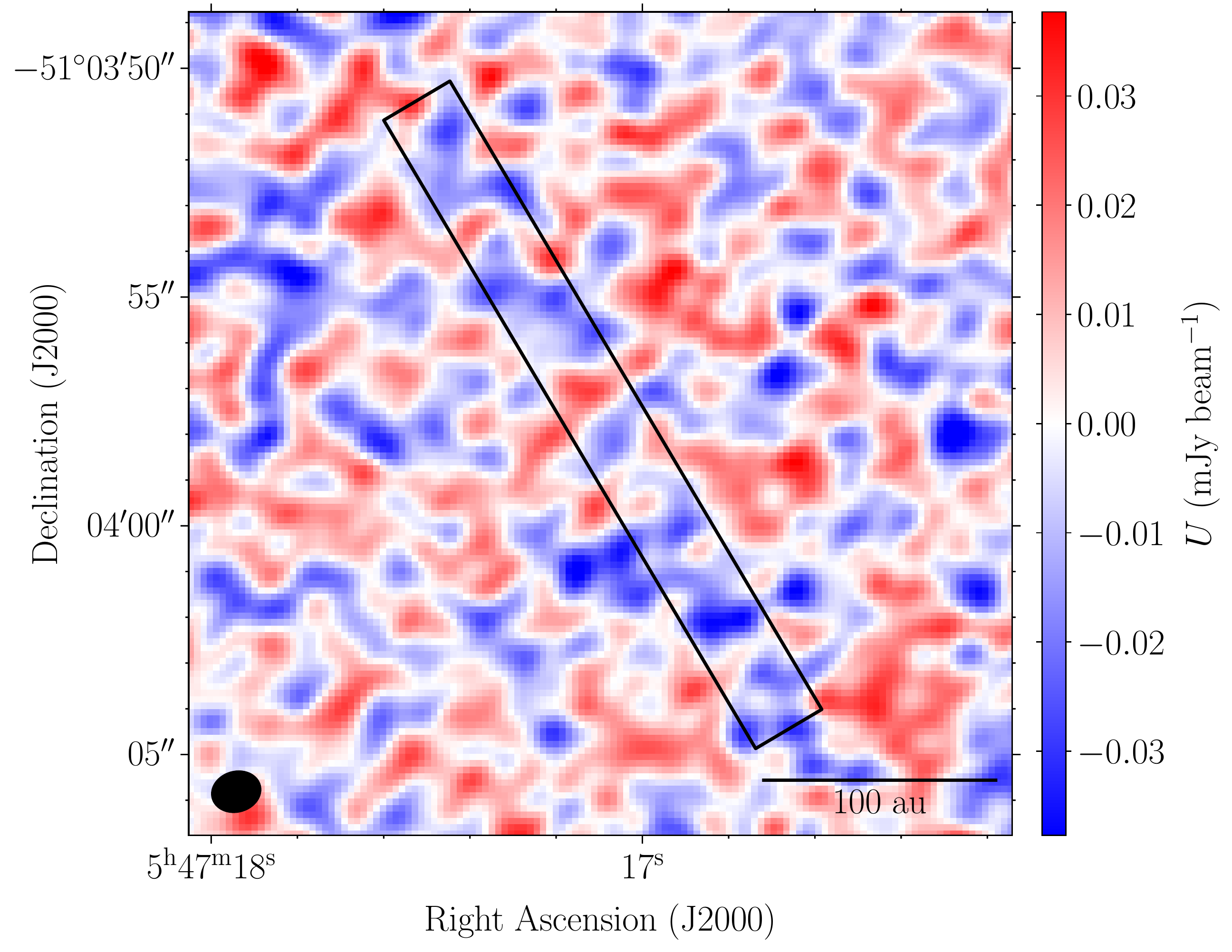}   
    \caption{
    870\,$\micron$ total intensity (Stokes $I$, top) and polarization (Stokes $Q$, center; and $U$, bottom) maps of the \betapic debris disk. 
    The Stokes $I$ map is dynamic-range limited, and is plotted beginning at $10\,\times$ the rms noise level of \rmsI\,\ujybm.  The peak of the Stokes $I$ emission is \peakI\,\mjybm.  The Stokes $I$ image has been corrected for the primary beam response; the integrated flux is \fluxI\,mJy.
    The $Q$ and $U$ maps are noise-like and are plotted between $\pm\,3\,\times$ the average rms noise level in the $Q$ and $U$ maps of \rmsQU\,\ujybm.  The $Q$ and $U$ maps have not been primary-beam corrected.
    The synthesized beam (resolution element) is shown as a black ellipse in the bottom-left corner of each panel and has dimensions of 1$\farcs$08\,$\times$\,0$\farcs$88 and a position angle of $-74.3\degree$.
    The rectangles in the $Q$ and $U$ images indicate the region where we average the $Q$ and $U$ emission (see Section \ref{sec:res}).
    (The data used to create this figure are available in the online version of this publication in AAS Journals.)
    }
    \label{fig:IQU}
\end{figure}

We use the task \texttt{TCLEAN} from CASA version 6.4.0.16 to produce images of \betapic, including of Stokes $I$, which corresponds to the total intensity dust emission, and of Stokes $Q$ and $U$, which correspond to linearly polarized emission.  We first make dirty images (i.e., with 0 clean iterations, corresponding to the source emission convolved with the ALMA point-spread function [PSF], or ``dirty beam'') of all three Stokes parameters.  It is at this step that we find that $Q$ and $U$ do not show any obvious emission; therefore, we do not clean $Q$ and $U$ further.  However, as Stokes $I$ exhibits ample signal, we clean the map using the \texttt{TCLEAN} auto-masking function (keyword: \texttt{auto-multithresh}) and performing 3225 iterations to clean down to a threshold of 16\,\ujybm.  Both the residual and final images of Stokes $I$ exhibit large-scale positive and negative ripples in the map at the $\pm$\,200\,\ujybm level (i.e., approximately $\pm$\,10\% of the peak Stokes $I$ level), suggesting that our observations are unable to recover large-scale emission in the field of view (single-dish observations have shown continuum emission extending along the major axis to scales greater than the $\sim$\,20$\arcsec$ scales probed by our ALMA observations; \citealt{Liseau2003}).  We are not able to reduce these features via self-calibrating the data, as the signal-to-noise ratio (SNR) of the Stokes $I$ emission is too low.  We thus proceed with the final images of the non-self-calibrated data despite the systematic ripples.  We should note, however, that these features do not affect our analysis, the majority of which hinges on the $Q$ and $U$ maps, which exhibit no such features and which achieve the expected thermal noise level of the observations.  The \texttt{robust}\,=\,2.0 maps that we analyze in the following sections are shown in Figure \ref{fig:IQU}.

The rms noise level in the dynamic-range-limited Stokes $I$ dust map is \rmsI\,\ujybm, calculated using the \texttt{sigma\_clipped\_stats} function from the \texttt{stats} module of the \texttt{astropy} Python package \citep{Astropy2018}.  The rms noise levels in the dirty $Q$ and $U$ images are \rmsQ\,\ujybm and \rmsU\,\ujybm, respectively, consistent with the thermal noise level expected given the on-source observation time.  

The Stokes $I$ flux density of \betapic derived using the primary-beam-corrected maps from our 2019 data is approximately \fluxI\,mJy.  This value is significantly lower than the 60\,mJy value quoted by \citet{Dent2014}; however, our value is consistent with the automated result from the Japanese Virtual Observatory (JVO)\footnote{JVO:  \url{http://jvo.nao.ac.jp/portal}} using the same data reported in \citeauthor{Dent2014}, and with the flux values in the images provided by the EA ARC when they delivered our data.  We thus assume that our derived flux value is correct.

The quantities (and upper limits) that can be derived from Stokes $I$, $Q$, and $U$ maps include the polarized intensity $P$, the linear polarization fraction $P_\textrm{frac}$, and the polarization position angle $\chi$ (measured E of N):

\begin{align}
P &= \sqrt{Q^2 + U^2} \label{eqn:P} \\
P_\textrm{frac} &= \frac{P}{I} \label{eqn:Pfrac} \\
\chi &= \frac{1}{2} \arctan{\left(\frac{U}{Q}\right)} \label{eqn:chi}\, .
\end{align}

\noindent
As $P$ is always a positive quantity (unlike the $Q$ and $U$ maps from which $P$ is derived, which can be either positive or negative), it has a positive bias. This bias is particularly significant in low-SNR measurements like those we report here.  We debias the polarized intensity values as described in \citet{Wardle1974,Hull2015b,Teague2021}.  Note that to debias a polarized intensity $P$ map one must use the rms noise value in the corresponding non-debiased $P$ map.  After debiasing the $P$ map, the rms noise value of the $P$ map becomes approximately equal to the noise values in the $Q$ and $U$ maps.  

The statistical uncertainties $\sigma_I$, $\sigma_Q$, $\sigma_U$, and $\sigma_P$ in the $I$, $Q$, $U$, and debiased $P$ maps are all equal to the rms noise values in the respective maps.  The uncertainty $\sigma_\chi$ in the polarization position angle $\chi$ is:

\begin{align}
\sigma_\chi &= \frac{1}{2} \frac{\sqrt{ (Q\sigma_U)^2 + (U\sigma_Q)^2 }}{P^2} \\
&\approx \frac{1}{2} \frac{\sigma_P}{P}\,\,.
\end{align}

\noindent
We make the simplification on the second line by assuming that $\sigma_Q \approx \sigma_U \approx \sigma_P$.  Note that this expression assumes a Gaussian distribution in position angles, which is not the case for low-SNR measurements \citep{Naghizadeh1993}.  We nevertheless proceed with these expressions to derive first-order estimates of the uncertainties in $\chi$, which are sufficient for our analysis.

The uncertainty $\sigma_{P_\textrm{frac}}$ in the polarization fraction $P_\textrm{frac}$ is:

\begin{align}
\sigma_{P_\textrm{frac}} &= P_\textrm{frac} \sqrt{ \left(\frac{\sigma_P}{P}\right)^2 + \left(\frac{\sigma_I}{I}\right)^2} \\
&\approx P_\textrm{frac} \frac{\sigma_P}{P} \\
&=\frac{\sigma_P}{I}\,\,,
\end{align}

\noindent
where $P$ is the debiased map of polarized intensity.  The simplification on the second line is appropriate for low-SNR data, where $(\sigma_P/P)^2$ ($= 0.25$ for $P = 2\,\sigma_P$) is much larger than $(\sigma_I/I)^2$ ($\approx 10^{-4}$ for a Stokes $I$ SNR of $\approx 100$ in the brightest regions of \betapic).

\section{Results}
\label{sec:res}

\begin{figure}
    \centering
    \includegraphics[width=0.48\textwidth, clip, trim=0cm 0cm 0cm 0cm]{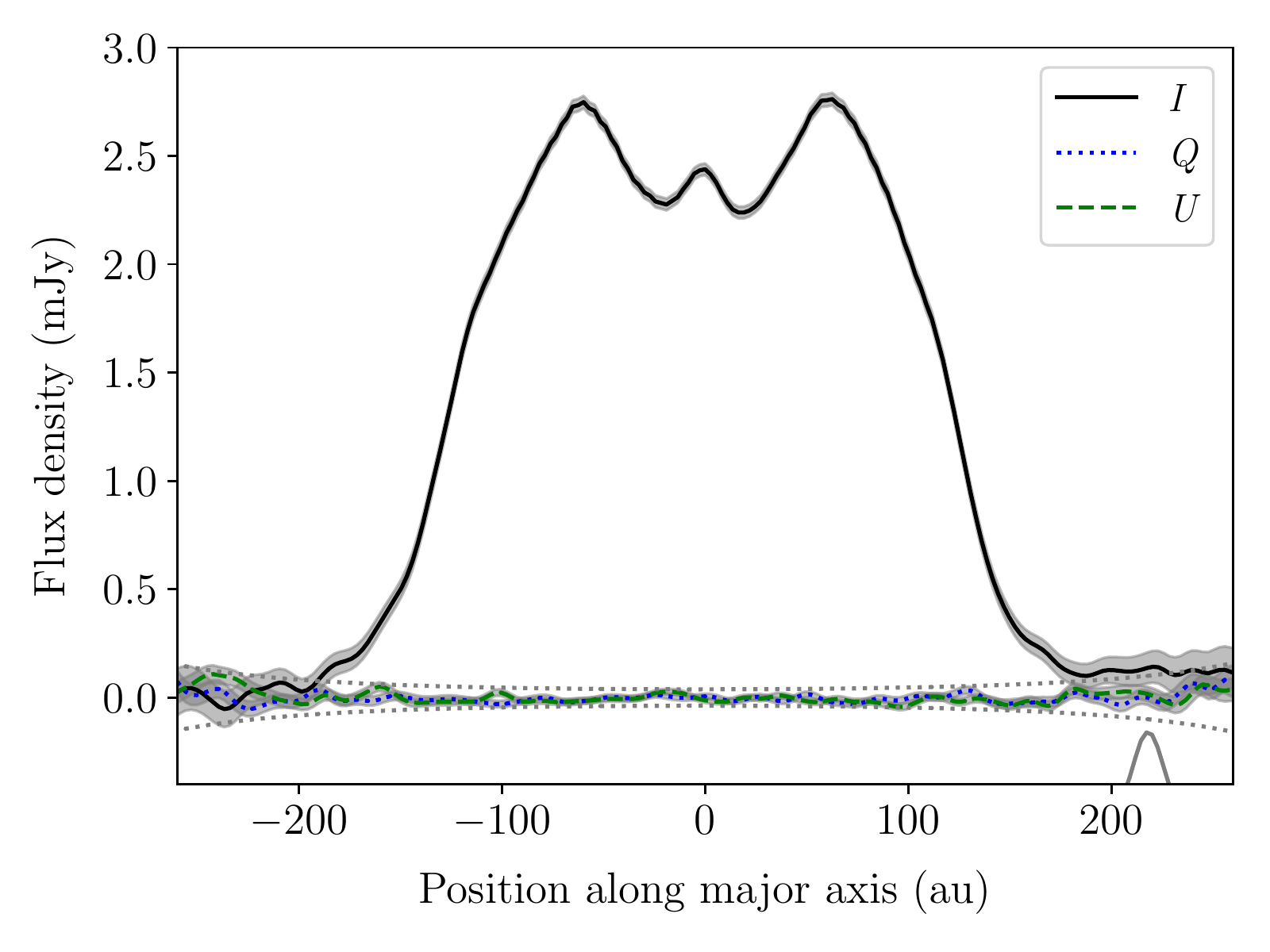}
    \caption{Cuts along the major axis of \betapic. The plot traces emission from the NE on the left (negative major axis values) to the SW on the right (positive values). The images used have been primary-beam corrected, and thus the $\pm$\,3\,$\sigma$ limits (dotted gray lines) are curved, increasing toward the edge of the image.  A cut through the $\sim$\,1$\arcsec$ synthesized beam (with the x-axis of the plot positioned at the beam's FWHM) is shown in the lower-right corner.}
    \label{fig:cuts}
\end{figure}

The $Q$ and $U$ images of \betapic (see the bottom two panels of Figure \ref{fig:IQU}) show emission consistent with noise.  We confirm this in Figure \ref{fig:cuts}, where we show $I$, $Q$, and $U$ cuts along the major axis of the disk. The cuts trace emission from the NE on the left (negative major axis values) to the SW on the right (positive values).  The $Q$ and $U$ profiles lie between the $\pm$\,3\,$\sigma$ curves (since $Q$ and $U$ can be positive or negative), where $\sigma$ is the average rms noise level in the $Q$ and $U$ maps.  The 3\,$\sigma$ curves in Figure \ref{fig:cuts} are curved as a result of the increase in the noise toward the edge of the maps, which (unlike the maps in Figure \ref{fig:IQU}) have been corrected for the primary-beam response of the small mosaic.

\begin{figure*}
    \centering
    \includegraphics[width=0.49\textwidth]{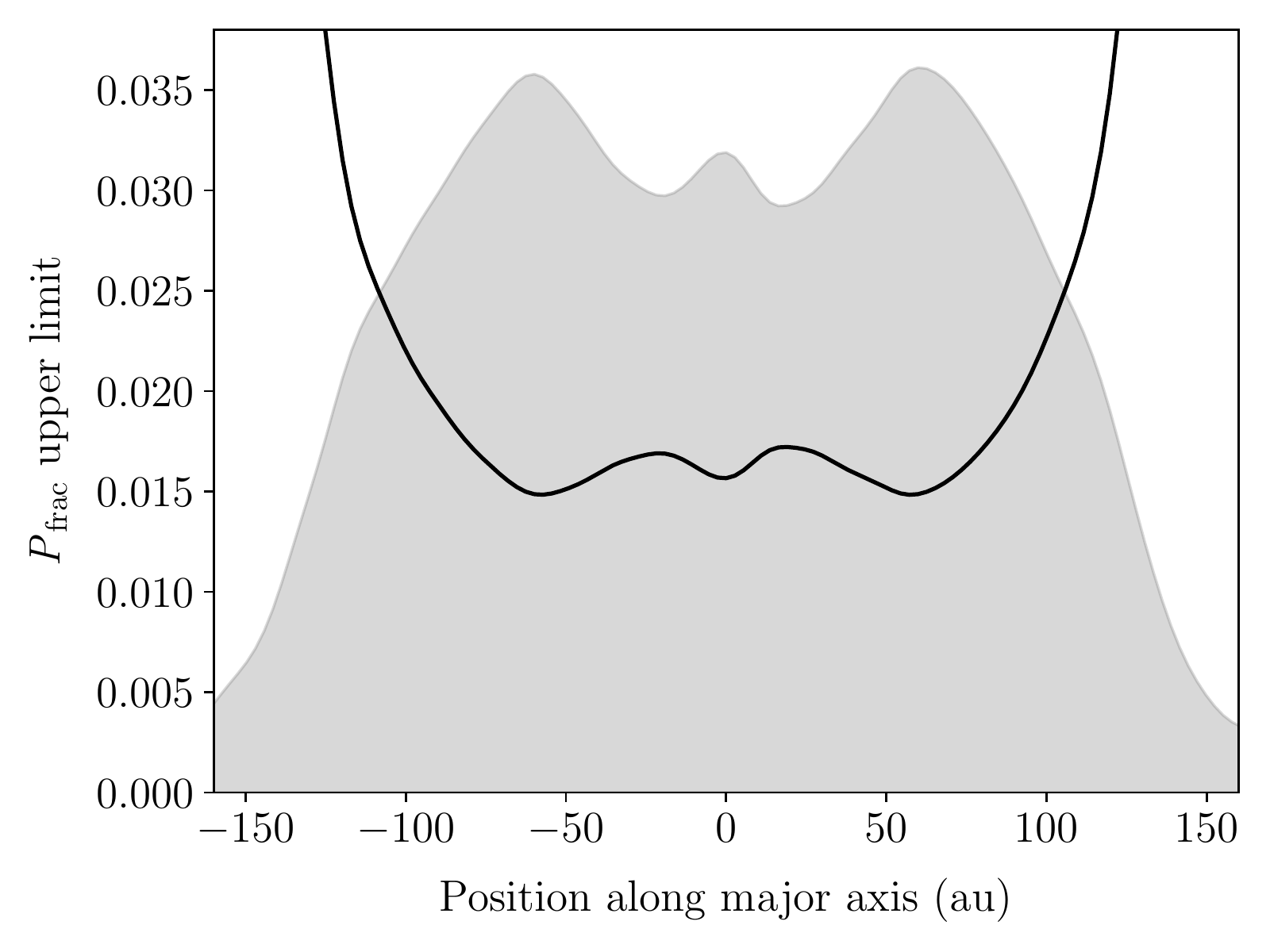}
    \includegraphics[width=0.49\textwidth]{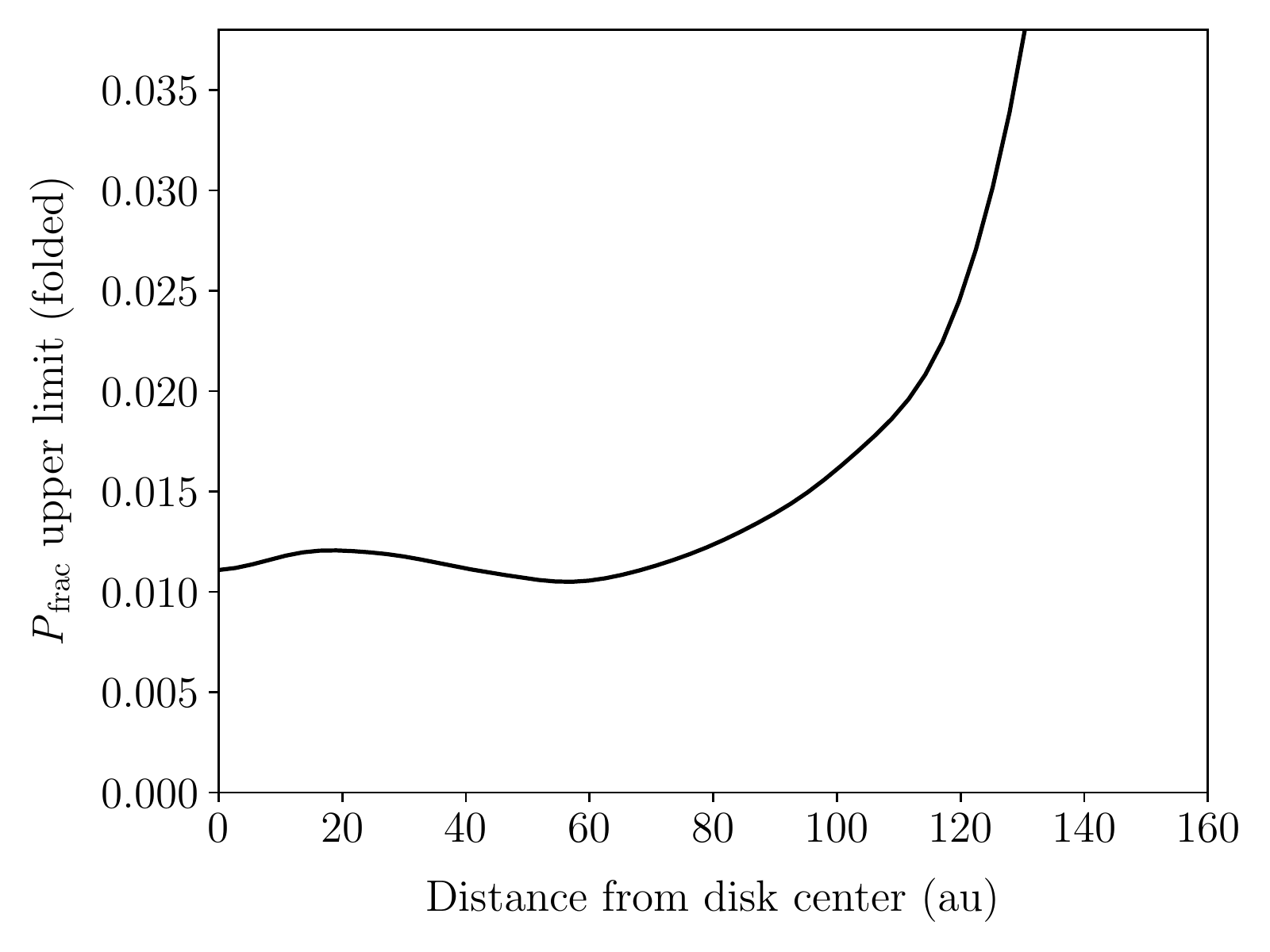}
    \caption{
    3\,$\sigma$ upper limits on the dust polarization fraction $P_\textrm{frac}$ in \betapic, where $\sigma$\,=\,\rmsQU\,\ujybm is the rms noise level in the spatially resolved, debiased $P$ map.  
    \textit{Left:} The solid line is the upper limit calculated by dividing 3\,$\sigma$ by the Stokes $I$ cut plotted in Figure \ref{fig:cuts}.
    The plot traces emission from the NE on the left (negative major axis values) to the SW on the right (positive values).
    The Stokes $I$ cut is shown in gray for reference, with arbitrary vertical units.
    \textit{Right:} 3\,$\sigma$ $P_\textrm{frac}$ upper limits plotted as a function of distance from the central star in \betapic.  We calculate the solid line by folding the Stokes $I$ data (i.e., averaging the data mirrored across the minor axis), and by using an rms noise level that is $\sqrt{2}$ lower than what was used in the left panel.
    Note that the upper limits are reported as fractions, not percentages.
    }
    \label{fig:pfrac_limits}
\end{figure*}

In Figure \ref{fig:pfrac_limits} we use the spatially resolved maps shown in Figure \ref{fig:IQU} to calculate upper limits on the polarization fraction $P_\textrm{frac}$ along the major axis of \betapic.  In the left panel we plot the 3\,$\sigma$ upper limits on $P_\textrm{frac}$, calculated by dividing 3\,$\times$ the off-source rms noise level in the debiased polarized intensity $P$ map (\rmsQU\,\ujybm) by the Stokes $I$ cut from Figure \ref{fig:cuts}.  In the right panel we plot the 3\,$\sigma$ upper limits on $P_\textrm{frac}$ after folding the Stokes $I$ data (i.e., averaging the data mirrored across the minor axis) and using an rms noise level in $P$ that is $\sqrt{2}$ lower than the value used in the left panel. The $P_\textrm{frac}$ upper limits within approximately $\pm$\,80\,au of the center of the \betapic debris disk are $\sim$\,\Pfraclimits\% (not folded) and $\sim$\,\Pfraclimitsfolded\% (folded). 

\subsection{$\sim$\,3\,$\sigma$ detection of dust polarization when averaging across the entire disk of \betapic}
\label{sec:detection}

To search for polarized dust emission below the noise level of the resolved maps, we average the emission across the entire disk of \betapic.  The main assumption that we make when performing this averaging is that the position angle of any polarized emission is the same everywhere: i.e., that it does not change as a function of position along either the minor or major axis of the disk.  This is reasonable given that, in an edge-on disk, the position angle of polarized dust emission should be uniformly along the minor axis of the disk in all of the following cases: scattering by dust grains \citep[e.g.,][]{CFLee2018a}; dust grains aligned via RATs with a toroidal magnetic field (known hereafter as $B$-RAT); and dust grains aligned with the (radial) radiation flux, or, more specifically, anisotropy in the radiation field (known hereafter as $k$-RAT; \citealt{LazarianHoang2007a,Tazaki2017}).  However, as we will see later, there is a key difference in the emission profiles from grains aligned via $B$-RAT versus $k$-RAT: in the case of $B$-RAT, the polarization fraction peaks near the center of the disk, whereas in the case of $k$-RAT the polarization fraction peaks near the outer regions of the disk (see Section \ref{sec:rt} and Figure \ref{fig:polarization}).

We average the maps using a box with dimensions of \boxminorpix$\,\times$\,\boxmajorpix\xspace pixels.  Each pixel in our maps is \pixel in size, yielding a box size of \boxminorarcsec\,$\times$\,\boxmajorarcsec, or \boxminorau\,au\,$\times$\,\boxmajorau\,au.  We perform this averaging on-source as well as in 14 off-source positions, seven on each side of the source.  The on-source averaging box covers the brightest part of the disk; we center the box at the position of the central star and set its size to maximize the SNR of the full-disk-averaged value of debiased $P$.  The off-source boxes are separated in the direction of the minor axis by 7 pixels; as the width of each box is \boxminorpix{} pixels, the sampling of the boxes in the map is slightly higher than Nyquist.

After computing full-disk-averaged values of $Q$ and $U$, we compute the polarized intensity $P$ (Equation~\ref{eqn:P}). In order to properly debias the full-disk-averaged $P$ value (see Section \ref{sec:obs}) and to calculate SNR values for $Q$, $U$, and debiased $P$, we need full-disk-averaged rms noise values, which we define to be the rms in the spatially resolved maps (\rmsP\,\ujybm for non-debiased $P$; \rmsQU\,\ujybm for $Q$ and $U$ and debiased $P$) divided by $\sqrt{\beamsInBox}$, where \beamsInBox{} is the number of synthesized beam areas contained in the averaging box.  The full-disk-averaged noise value for non-debiased $P$ is \rmsPboxavg\,\ujybm; for $Q$ and $U$ (and debiased $P$) the value is \rmsQUboxavg\,\ujybm.  We use the former noise value to debias the full-disk-averaged $P$ value and the latter value to calculate SNR values.  We also calculate the polarization fraction $P_\textrm{frac}$ (Equation~\ref{eqn:Pfrac}, using the full-disk-averaged $I$ and debiased $P$ values) and position angle $\chi$ (Equation~\ref{eqn:chi}).  

\begin{deluxetable}{cc}
\tabletypesize{\normalsize}
\tablecaption{Full-disk-averaged data}
\tablehead{
\colhead{Quantity} & \colhead{Value} \\ [-1.8em]
\colhead{}         & \colhead{}
} 
\startdata \\ [-1.2em]
$I$ & \Iboxavg\ \ujybm \\ [0.8em]
$Q$ & \Qboxavg\ \ujybm (SNR = \SNRQboxavg) \\
$U$ & \Uboxavg\ \ujybm (SNR = \SNRUboxavg) \\
$\chi$ & \PAboxavg$\degree$ \\ [0.8em]
$Q^\prime$ & \Qprimeboxavg\ \ujybm (SNR = \SNRQprimeboxavg) \\
$U^\prime$ & \Uprimeboxavg\ \ujybm (SNR = \SNRUprimeboxavg) \\
$\chi^\prime$ & \PAprimeboxavg$\degree$ \\ [0.8em]
$P$ & \Pboxavg\ \ujybm (SNR = \SNRPboxavg) \\
$P_\textrm{frac}$ & \Pfracpctboxavg\% \\ [0.3em]
\enddata
\tablecomments{
All flux density values have been averaged across the brightest part of the \betapic debris disk, as described in Section \ref{sec:detection}.  $Q^\prime$, $U^\prime$, and $\chi^\prime$ reflect the values after performing the transformation described in Equations \eqref{eq:transformQ} and \eqref{eq:transformU}.  The value of the polarized intensity $P$ is debiased as described in Section \ref{sec:obs}. All SNR values are calculated using the rms noise value of \rmsQUboxavg\,\ujybm from the full-box-averaged $Q$, $U$, and debiased $P$ maps (see Section \ref{sec:detection}).  For reference, the position angle of the minor axis of the disk is $-59\degree$ in the original reference frame and $0\degree$ in the rotated reference frame.
}
\label{table:data}
\end{deluxetable}

\begin{deluxetable}{ll}
\tabletypesize{\normalsize}
\tablecaption{RMS noise values}
\tablehead{
\colhead{Noise Value} & \colhead{Description} \\ [-0.3em]
\colhead{(\ujybm)}         & \colhead{}
} 
\startdata \\ [-1.2em]
\rmsI & Stokes $I$ noise \\
\rmsQU & Mean of Stokes $Q$ and $U$ noise \\ 
\phantom{0}\rmsQUboxavg & Mean of full-disk-avg.\ $Q$ and $U$ noise \\ [0.2em]
\enddata
\tablecomments{
Relevant noise values.  
Rows 1--2 are from the spatially resolved maps (Section \ref{sec:res}).  
Row 3 is from the full-disk-averaged maps (Section \ref{sec:detection}).
After debiasing $P$ (whether spatially resolved or full-disk averaged), the noise in the $P$ map is approximately equal to the noise in the associated $Q$ and $U$ maps.
}
\label{table:noise}
\end{deluxetable}

The on-source results are as follows. The SNR of the debiased $P$ value, defined as $P$ divided by the full-disk-averaged rms 
noise, is \SNRPboxavg, with corresponding $Q$ and $U$ SNR values of \SNRQboxavg and \SNRUboxavg, respectively. The polarization fraction $P_\textrm{frac}$\,=\,\Pfracboxavg (i.e., \Pfracpctboxavg\%) and polarization position angle $\chi$\,=\,\PAboxavg$\degree$.  This value of $\chi$ matches the \minoraxis$\degree$ position angle of minor axis of \betapic to within the uncertainty of the position angle.  We perform the same analysis for all of the 14 off-axis positions.  The SNR value of the debiased $P$ is by far the highest (with an SNR of \SNRPboxavg) in the on-axis position.  Two off-axis positions have debiased $P$ SNR values of 1.9 and 0.8; the remainder of the positions have SNR values of 0.

Finally, in an attempt to achieve a higher-SNR detection from the $Q$ and $U$ maps, we transform the $Q$ and $U$ data into the $Q^\prime$ and $U^\prime$ frame as follows:

\begin{align}
    Q^{\prime} &= Q \cos(2\theta) + U \sin(2\theta)
    \label{eq:transformQ}\\
    U^{\prime} &= U \cos(2\theta) - Q \sin(2\theta)\,\,,
    \label{eq:transformU}
\end{align}

\noindent
where $\theta = 90\degree - $\majoraxis$\degree$; the latter angle (\majoraxis$\degree$) is the orientation of the major axis of the disk.  In the transformed ($Q^\prime$,\,$U^\prime$) reference frame, $+Q^\prime$ is oriented along the minor axis of the disk. This transformation is similar to the one used by \citet{Schmid2006, Teague2021}.  However, whereas those authors used this method to transform centro-symmetric (e.g., radial or azimuthal) polarization patterns as a function of polar angle on the sky, our application is simpler: we apply the same transformation to each pixel in our images, since the polarization from \betapic is assumed to have the same orientation everywhere in the disk.

Assuming, as we have thus far, that all polarized emission from \betapic should have a position angle along the minor axis of the disk, we would expect the transformed data to exhibit substantial positive signal in $Q^\prime$. This is indeed what we find.  While the debiased $P$ value of the transformed data has an identical SNR of \SNRPboxavg (as expected), we find that $Q^\prime$ is positive and has a statistically significant SNR of \SNRQprimeboxavg.  $U^\prime$ is consistent with noise, and the corresponding position angle $\chi^\prime$ is consistent with 0$\degree$ (i.e., along the minor axis of the disk in the transformed reference frame). See Table \ref{table:data} for a summary of our observational results and Table \ref{table:noise} for a list of noise values relevant to our analysis.

\subsection{Polarized dust emission in the middle versus outer regions of \betapic}
\label{sec:divided}

Considering the significant detection of polarized emission when averaging across the entire disk, the final tests we perform are to search for polarized emission in the inner region of the disk (where we would expect to see more polarized emission from grains aligned via $B$-RAT) versus the outer regions (where polarized emission from grains aligned via $k$-RAT should dominate).  We perform our averaging tests using different sections (e.g., quarters, thirds, halves) of a box that is slightly longer than the one used for the full-disk averaging described above. The box has dimensions of \boxdividedminorpix$\,\times$\,\boxdividedmajorpix\xspace pixels, corresponding to a box size of \boxdividedminorarcsec\,$\times$\,\boxdividedmajorarcsec{} or \boxdividedminorau\,au\,$\times$\,\boxdividedmajorau\,au.

Our first test is to analyze the middle half of the disk versus the (combined) outer two quarters. We do not detect significant polarization in the inner half of the disk centered on the central star of \betapic.  However, the combined outer two quarters exhibit polarized emission at the \Pfracouter\% level with a marginally significant SNR in the debiased $P$ map of \SNRPouter (the SNR in the $Q^\prime$ map is \SNRQpouter).  The fact that the polarized emission is only detected in the outer regions of \betapic suggests that dust grains aligned via $k$-RAT are producing the polarized emission, as we will discuss later in more detail.

Given that the polarized emission appears to be coming from the outer regions of the disk, our second test is to analyze the NE third, center third, and SW third of the disk separately.  We find no detectable polarization in the NE or center thirds; we only detect polarized emission in the SW third in the debiased $P$ map at the \PfracSWthird\% level with a marginally significant SNR of \SNRPSWthird (the SNR in the $Q^\prime$ map is \SNRQpSWthird).  This asymmetry in the polarized emission is unexpected given the symmetry of the Stokes $I$ emission in our 870\,$\micron$ ALMA observations (Figure \ref{fig:IQU}, top panel).  However, asymmetries in the Stokes $I$ dust emission are seen at other wavelengths, from the optical \citep{Apai2015} to the mid-infrared \citep{Telesco2005} to 1.3\,mm observations by ALMA; the latter suggest that the dust peak on the SW side of the disk is $\sim$\,13\% brighter than the NE peak.\footnote{This can be seen in JVO images of 1.3\,mm dust continuum observations of \betapic from ALMA project 2018.1.00072.S.}  It is unclear whether the asymmetry seem at other wavelengths is related to the asymmetry in the polarized emission.  We leave the topic of asymmetry for future studies, and proceed to interpret our observations using symmetric models.  In the sections that follow we endeavor to explain the low level of dust polarization in \betapic by exploring the parameter space of our theoretical models of dust-grain alignment via RATs in a debris disk.

\section{Model of the \betapic debris disk}
\label{sec:model}

\begin{figure}
    \centering
    \includegraphics[width=0.48\textwidth, clip, trim=0.5cm 0cm 0cm 0cm]{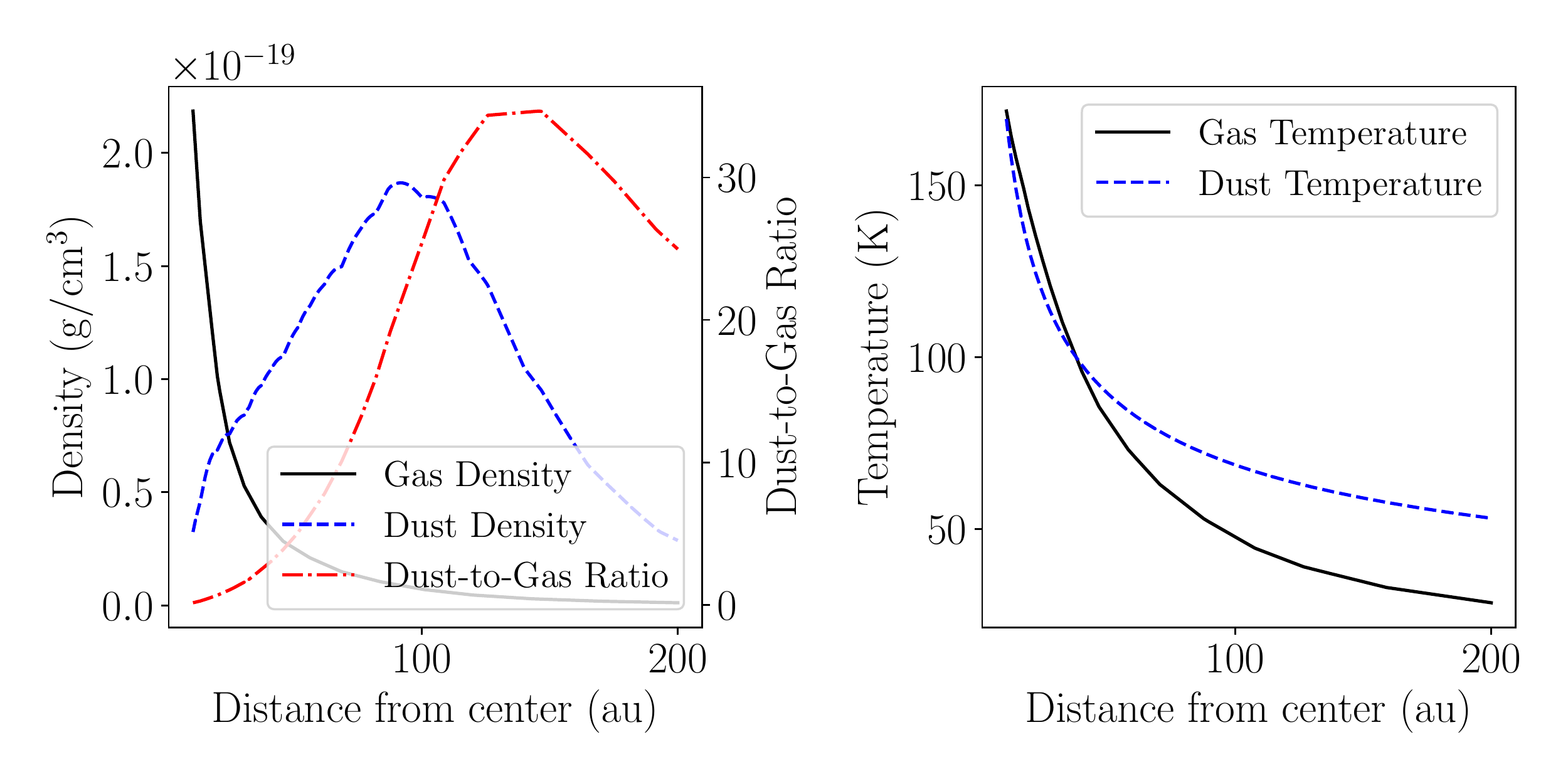}
    \caption{Various radial profiles in our model, based on \citet{Zagorovsky2010, Kral2016}. \textit{Left}: Gas mass density, dust mass density, and dust-to-gas mass ratio. \textit{Right}: Gas and dust temperatures.}
    \label{fig:model}
\end{figure}

We use the \betapic debris disk model presented in \citet[][hereafter \citetalias{Kral2016}; see their Figure 9]{Kral2016}. In this model, the gaseous component of the disk is composed of carbon and oxygen atoms with equal number densities (unlike typical protoplanetary disks that are dominated by molecular hydrogen, \betapic's gaseous component is of secondary origin, and the molecular hydrogen density is small: see, e.g., \citealt{Matra2017}). We derive the gas mass density $\rho_g$ and the gas number density $n_g$ from the \ion{O}{1} number density. The dust density $\rho_d$ is taken from \citet{Zagorovsky2010}, who determine it empirically by fitting scattered-light dust observations of \betapic from the Hubble Space Telescope/STIS \citep{Heap2000}.\footnote{The large, millimeter-sized grains seen by ALMA are located between 50-130\,au \citep{Dent2014}.  In contrast, the smaller, micron-sized grains seen in scattered light and in the mid-infrared extend from <\,30\,au (originating in a possible inner disk: \citealt{DLi2012, Apai2015, MillarBlanchaer2015}) to beyond 2000\,au (because of radiation pressure).} Together, the values of $\rho_d$ and $\rho_g$ yield a dust-to-gas mass ratio of up to approximately 30:1, far higher than the typically assumed ratio of 1:100 in protoplanetary disks and in the galactic interstellar medium \citep{Bohlin1978}. 
$\rho_d$, $\rho_g$, and the dust-to-gas ratio are plotted in the left panel of Figure~\ref{fig:model}.

The gas temperature $T_g$ and the gas scale height are taken directly from \citetalias{Kral2016}, where they are calculated self-consistently via a photodissociation-region model. 
For simplicity, we assume the dust scale height is the same as that of the gas.\footnote{The dust in \betapic most likely resides in more complicated structures than those captured by our simple model. \cite{Matra2019} modeled the vertical distribution of dust using ALMA Band~6 (1.3\,mm) continuum emission and found that there are two distributions of dust with different scale heights of $5.1$\,au and $15.7$\,au. For comparison, the scale height in our model is $\sim5$\,au at a radius of $100$\,au. However, because our spatial resolution corresponds to roughly $19$\,au, larger than both of the scale heights from \cite{Matra2019}, the vertical structure is unresolved and thus has no effect on our results. \label{footnote:scaleheight} } 
We calculate the dust temperature using the Monte Carlo Radiative Transfer code RADMC-3D \citep{Dullemond2012} assuming a grain size of $13.8\rm\, \mu m$.
We also calculate the temperature assuming large grains with a size of $1.38\rm\, mm$ and find that the dust temperature can differ by $\sim$\,20\% at $100$\,au with respect to the small-grain case. However, this difference has little effect on the polarization fraction profile, which is the main focus of this paper that allows us to distinguish different grain-alignment mechanisms. The impact of the dust temperature on the intensity is compensated by a universal density scaling factor (see Section~\ref{sec:synobs}). The dust temperature also affects the timescales relevant for grain alignment, but the $20\%$ difference is inconsequential (see Section~\ref{sec:theory}). 
The gas and dust temperature distributions are plotted in the right panel of Figure~\ref{fig:model}. 

Using the above values, we calculate the column density of dust for an edge-on view of the disk. We find that the column density peaks around $75$\,au at a value of $8.2\times 10^{-4}\,\rm g\,cm^{-2}$. 
We thus find that the \betapic debris disk is very optically thin, unless the dust opacity is many orders of magnitude higher than $1\rm\, cm^2/g$, which is unrealistic. The scattering optical depth is also expected to be much less than unity. We thus ignore scattering-induced dust polarization and consider only polarized thermal emission from aligned dust grains.\footnote{This is further justified by our synthetic observations (see Section~\ref{sec:synobs}): when grain alignment is turned off, the scattering-induced polarization fraction is typically on the order of $10^{-7}$. We do not discuss this result further in this paper.}

The above information is sufficient to allow us to conduct radiative transfer calculations in Sections~\ref{sec:rt} and \ref{sec:synobs}. Additional information about the strength of the magnetic and radiation fields is needed to analyze the alignment of dust grains, which we discuss in detail in Section~\ref{sec:theory}.

\section{Constraints on the intrinsic dust polarization fraction}
\label{sec:rt}

Here we present a simple semi-analytical radiative transfer model whose parameter space we can explore quickly. The results have been checked against the Monte Carlo radiative transfer calculations that we present in Section \ref{sec:synobs} and show good agreement. In this model we assume an axisymmetric disk with the density and temperature profiles prescribed in Section \ref{sec:model}.

\subsection{Dust model}
For simplicity and ease of computational cost, we assume small dust grains in the dipole regime.
For the alignment of dust grains, we focus on the RAT mechanism. There are two possible configurations: grains are aligned either with the (radial) radiation flux ($k$-RAT) or with the magnetic field ($B$-RAT).  Hereafter, we will sometimes refer to the radiation and magnetic fields as the ``aligning fields.''
When RAT is operating, the dust grains are aligned with their short axes along the aligning field and produce polarization perpendicular to it. As such, dust grains can be well represented by oblate spheroids (see, e.g., \citealt{Yang2019}).
In the dipole regime, we assume that these small, oblate-spheroidal dust grains have polarizability $\alpha_1$ and $\alpha_3$ along the long and short axes of the dust grain, respectively \citep{BH83,Yang2016b}. The absorption cross section of the dust grains is:
\begin{equation}
\sigma_\textrm{abs} = k\,\mathrm{Im}[\alpha_1 (1+\cos^2 i) + \alpha_3 \sin^2 i]
\label{eq:sig_abs}
\end{equation}
and the polarization cross section is:
\begin{equation}
\sigma_{p} = k\,\mathrm{Im}[(\alpha_1- \alpha_3) \sin^2 i]\,\,,
\label{eq:sig_p}
\end{equation}
where $i$ is the angle between the symmetry axis of the dust grain and the light propagation direction, and $k=2\pi/\lambda$ is the wave number.
With this dust model, the polarization fraction at $i=\pi/2$ is:
\begin{equation}
p_0 = \frac{\sigma_p}{\sigma_\textrm{abs}} = \frac{\mathrm{Im}[\alpha_1-\alpha_3]}{\mathrm{Im}[\alpha_1+\alpha_3]}\,\,.
\label{eq:p_0}
\end{equation}
We will refer to $p_0$ as the ``intrinsic polarization fraction,'' which is the polarization fraction of the thermal dust emission when the dust grains are in an optically thin medium and are all uniformly viewed as edge-on by the observer.
According to this model, the polarization fraction as a function of inclination angle is:
\begin{equation}
\begin{split}
p(i)&=\frac{\mathrm{Im}[(\alpha_1- \alpha_3) \sin^2 i]}{\mathrm{Im}[\alpha_1 (1+\cos^2 i) + \alpha_3 \sin^2 i]}\\
&= p_0\sin^2 i + O(p_0^2)\,\,.
\end{split}
\label{eq:pprof}
\end{equation}
Equation~\eqref{eq:pprof} shows that $p_0$ and the geometric factor $\sin^2 i$ can be separated completely to the leading order.
As such, the observed polarization fraction scales approximately linearly with the intrinsic polarization fraction $p_0$ for any given geometry of the aligning fields along the line of sight.

The observed polarization depends not only on $p_0$, but also on the geometry of the underlying aligning field along the line of sight and on the degree of grain alignment. The geometric effect will be modeled later. Here we discuss the observable effect of the degree of grain alignment. 
If the dust grains are perfectly aligned, the polarization profile is well described by Equation~\eqref{eq:pprof}.
If the dust grains are poorly aligned, the polarization profile will also follow Equation~\eqref{eq:pprof}, except that $p_0$ will be replaced with $Rp_0$, where $R$ is the so-called ``Rayleigh reduction factor'' \citep{Lee1985}:
\begin{equation}
R = \frac{3}{2}\left(\left<\cos^2\eta\right>-\frac{1}{3}\right),
\end{equation}
where $\left<\cos^2\eta\right>$ is averaged over the ensemble of dust grains, with $\eta$  being the angle between the symmetry axis of dust grains and the alignment axis. For perfectly aligned grains, $\left<\cos^2\eta\right>=1$ and $R=1$. For non-aligned grains, $\left<\cos^2\eta\right>=1/3$ and $R=0$. 
Because $R$ and $p_0$ are always multiplied by one another, we cannot tell them apart observationally. In the following radiative transfer calculations and in Section~\ref{sec:synobs}, we will assume $R=1$ and take the intrinsic polarization fraction $p_0$ as the single parameter describing our dust grains.

\subsection{Polarization profiles for $k$-RAT and $B$-RAT}
\begin{figure}
    \centering
    \includegraphics[width=0.45\textwidth]{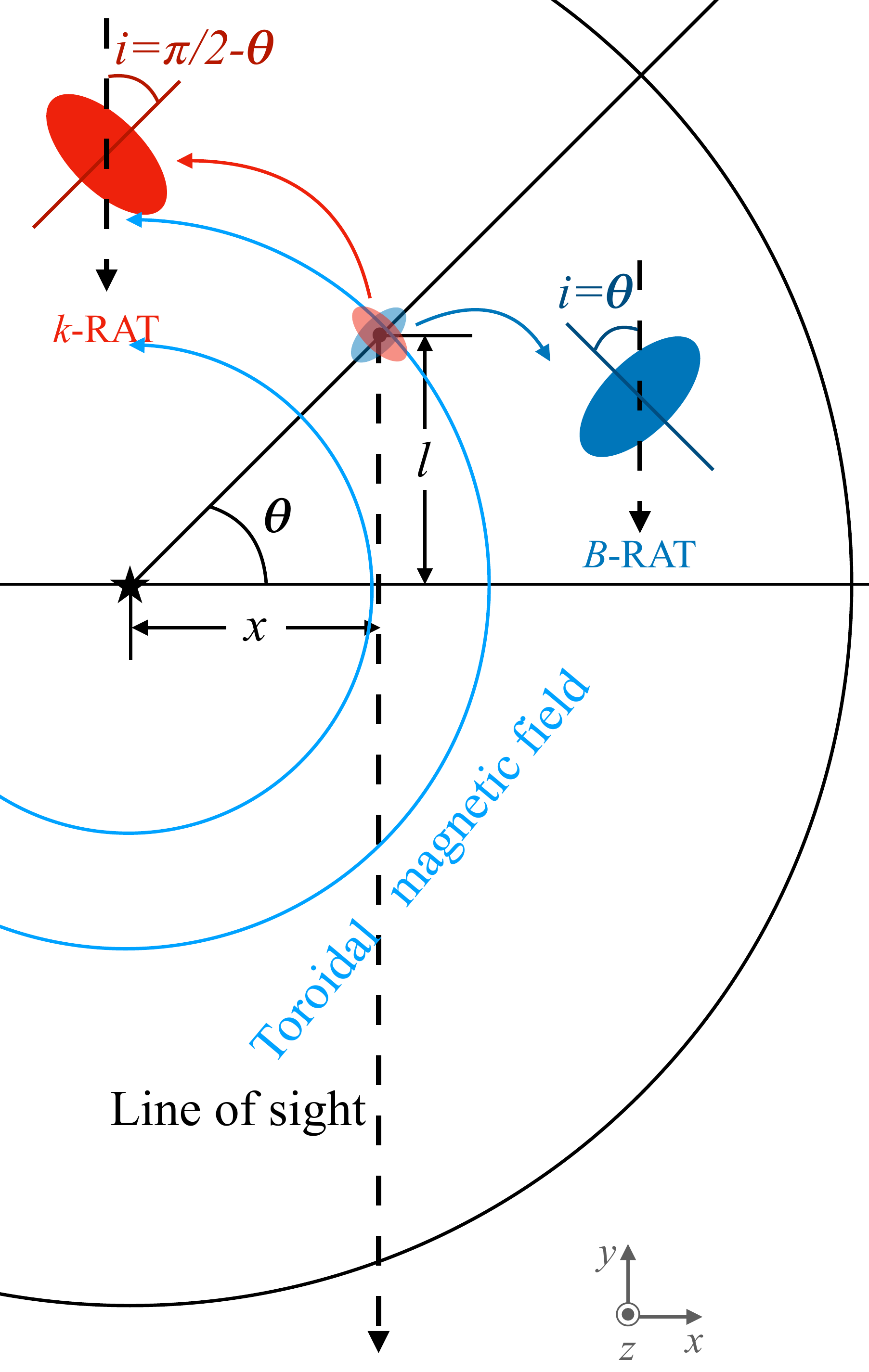}
    \caption{The setting of our semi-analytical radiative transfer model. The black arc represents the \betapic debris disk. The horizontal line is the sky plane, which is perpendicular to our line of sight, shown as a dashed arrow. A representative dust grain (black dot) has an azimuth angle $\theta$ measured counterclockwise from the sky plane. $x$ is the distance from the line of sight to the center of the disk. $l$ is the distance along the line of sight, with $l=0$ being in the plane of the sky. We consider two alignment mechanisms. For $k$-RAT, we have a dust grain (red oval) whose short axis is aligned with the radial radiation flux, which thus makes an angle $i=\pi/2-\theta$ with the line of sight. For $B$-RAT, we have a dust grain (blue oval) whose short axis is aligned with a toroidal magnetic field such that the symmetry axis of the dust grain makes an angle $i=\theta$ with the line of sight.}
    \label{fig:rtsetting}
\end{figure}

In the $k$-RAT regime, the dust grains are aligned with their short axes along the direction of the radiation flux. Given our prescribed axisymmetric model, the radiation flux can only be in the radial direction. The dust grains aligned with such a radiation field are thus oriented with their short axes along the radial direction. 
For a dust grain placed in the debris disk with azimuth angle 
$\theta$, the angle between the line of sight and the symmetry axis of the dust grain is $i = \pi/2-\theta$. 
The Stokes $I$ (total intensity) dust emission from $k$-RAT, $I_k$, can then be calculated as:
\begin{equation}
\begin{split}
I_k \propto &\int_{-\infty}^\infty \rho(r(x,l))\,T(r(x,l))\\
&\left[\omega(1+\sin^2\theta(x,l)) +\cos^2\theta(x,l)\right] dl\,\,,
\end{split}
\end{equation}
where $\rho(r)$ is the density profile and $T(r)$ is the temperature profile. 
$r(x,l)=\sqrt{x^2+l^2}$ and $\theta(x,l)=\tan^{-1}(l/x)$ are the distance from the center and the azimuthal angle in the disk,
respectively, where $x$ is the location of the
line of sight in the sky plane and $l$ is the distance along the line of sight ($l=0$ lies along the $x$-axis). 
$\omega\equiv\mathrm{Im}[\alpha_3]/\mathrm{Im}[\alpha_1]$ is the ratio of the absorption cross sections along the two principle axes, which is related to the intrinsic polarization fraction as $\omega=(1+p_0)/(1-p_0)$.
We illustrate the geometry of our setting in Figure~\ref{fig:rtsetting}, where the red oval represents a grain aligned via the $k$-RAT mechanism.

We define the Stokes parameters in our models such that $Q>0$ implies polarization along the $z$ axis (see Figure \ref{fig:rtsetting}), i.e., the direction perpendicular to the plane of the debris disk:
\begin{equation}
\begin{split}
Q_{k} \propto &\int_{-\infty}^\infty \rho(r(x,l))\,T(r(x,l)) \\ &(\omega-1)\cos^2\theta(x,l) \,dl\,\,.
\end{split}
\end{equation}

\noindent
This definition of $Q$ in our models is analogous to our definition of $Q^\prime$ in the transformed reference frame discussed in Section \ref{sec:detection}. 

For simplicity, in the $B$-RAT regime we assume a purely toroidal magnetic field. A toroidal field configuration is a natural outcome in both a rotating disk and in the outer reaches ($\sim$\,50--100\,au in the case of the Sun) of the stellar-dominated heliosphere of a rotating star \citep{Owens2013}, and thus should be the magnetic field configuration near the disk midplane where most dust grains reside.
While observations of protoplanetary disks suggest that some may have significant poloidal components of their magnetic fields \citep{DLi2016,Alves2018}, a predominantly poloidal magnetic field in \betapic would produce polarization along the major axis of the debris disk, which is perpendicular to the full-disk-averaged polarization orientation that we observe. Thus, if $B$-RAT is the cause of the polarization signal, poloidal magnetic fields are most likely playing a subdominant role in the \betapic system.\footnote{Note that we cannot rule out poloidal magnetic fields completely here. While small grains aligned with poloidal magnetic fields will produce polarization along the major axis of the disk, which is the opposite of what we see, larger grains that are aligned via $B$-RAT can experience the effect of polarization reversal, or ``negative polarization'' (see, e.g., \citealt{Guillet2020}), which produces polarization along the magnetic field direction instead of perpendicular to it. Such complications are left for future studies.}
In this regime, dust grains are aligned with their short axes along the magnetic field direction, with the symmetry axis of the oblate dust grain making an angle of $i=\theta$ with the line of sight. We represent the grain aligned via $B$-RAT as a blue oval in Figure~\ref{fig:rtsetting}.
The Stokes $I$ emission from $B$-RAT, $I_B$, can then be calculated as:
\begin{equation}
\begin{split}
I_B \propto &\int_{-\infty}^\infty \rho(r(x,l))\,T(r(x,l))\\
&\left[a(1+\cos^2\theta(x,l)) +\sin^2\theta(x,l)\right] dl\,\,.
\end{split}
\end{equation}
With the same definition of Stokes $Q$, we have:
\begin{equation}
\begin{split}
Q_{B} \propto &\int_{-\infty}^\infty \rho(r(x,l)) T(r(x,l))\\ &(\omega-1)\sin^2\theta(x,l) \,dl\,\,.
\end{split}
\end{equation}

\begin{figure}
    \centering
    \includegraphics[width=0.48\textwidth]{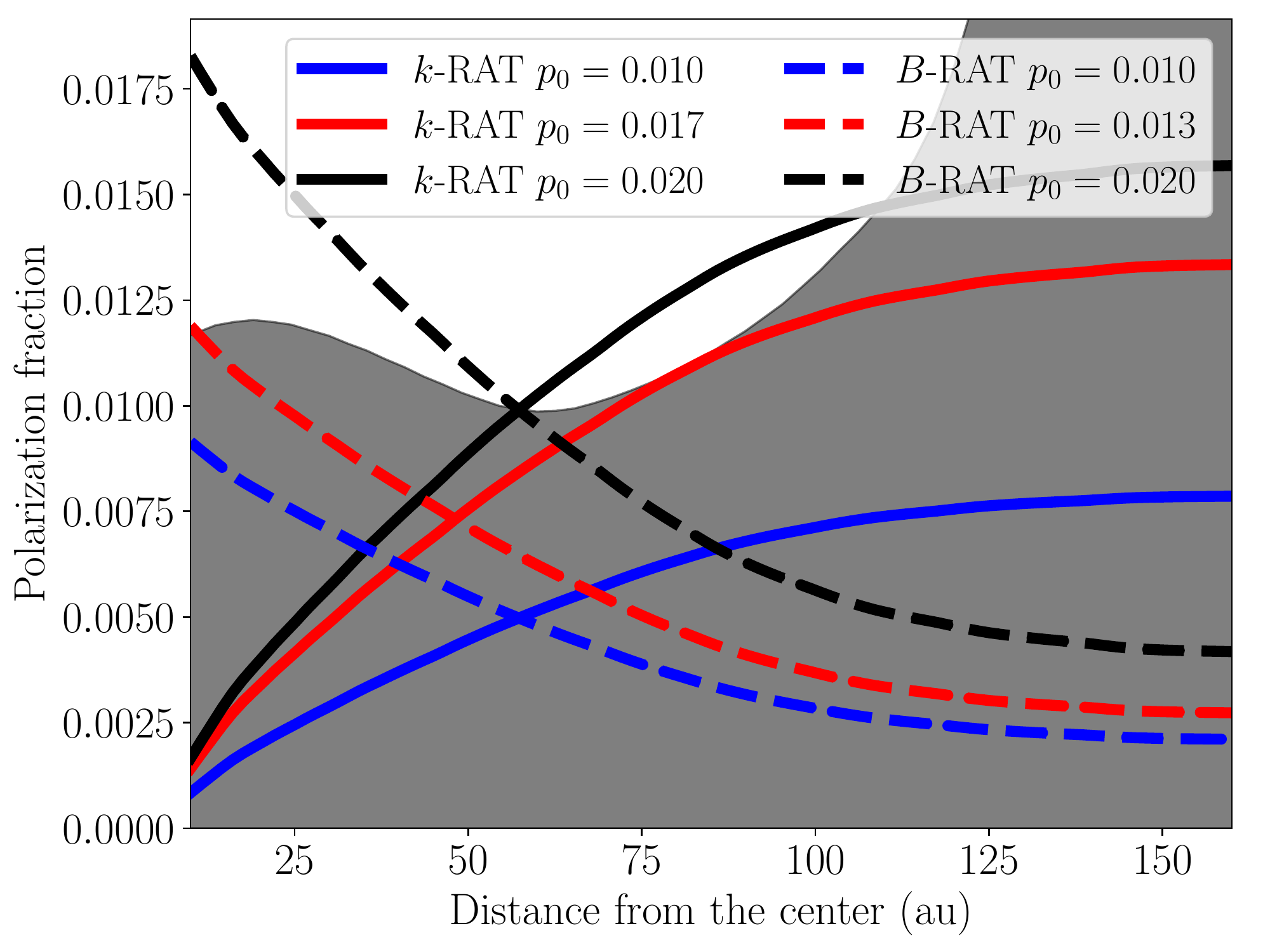}
    \caption{Polarization fraction as a function of distance from the center of our modeled debris disk. Solid lines show results for $k$-RAT alignment. Dashed lines show results for $B$-RAT alignment. Different colors represent dust grains with different values of the intrinsic polarization fraction $p_0$. The shaded region is the region permitted by our ALMA observations; the upper envelope of the region is the same as the curve shown in the right-hand panel of Figure \ref{fig:pfrac_limits}. }
    \label{fig:polarization}
\end{figure}

We can now calculate the polarization fraction as a function of $x$, the distance from the center of the disk, for grains aligned via both the $k$-RAT and $B$-RAT mechanisms. We plot the results in Figure~\ref{fig:polarization}.  While both $k$-RAT and $B$-RAT predict polarization angles aligned with the minor axis of the disk, they have different polarized intensity profiles as a function of distance along the major axis.  We can see that for $k$-RAT (solid curves), polarization increases toward larger radii. The strongest constraint on our $k$-RAT models comes from the non-detection between 70--100\,au in the folded ALMA data (see the right panel of Figure \ref{fig:pfrac_limits}), which allows us to exclude models with intrinsic polarization fractions $p_0\gtrsim 1.7\%$. On the other hand, if we assume $B$-RAT (dotted curves), the polarization decreases toward larger radii. The strongest constraint on our $B$-RAT models comes from the observational upper limits toward the center of the disk, where we expect a large amount of polarization. In this case, we can put a slightly stronger constraint on the intrinsic polarization fraction, excluding models with $p_0\gtrsim1.3\%$. 

We can see that the difference in polarization profile is the key to distinguishing between $k$-RAT and $B$-RAT. These different profiles  are the result of the geometry of dust grains (as viewed by the observer) that have been aligned with respect to the different aligning fields: the radiation field in the case of $k$-RAT, and the magnetic field in the case of $B$-RAT. For $k$-RAT ($B$-RAT), dust grains are aligned with radial (toroidal) fields, which are parallel (perpendicular) to the line of sight near the center of the disk and perpendicular (parallel) to the line of sight toward the edges of the disk. Hence, dust grains aligned via $k$-RAT ($B$-RAT) are viewed by the observer to be face-on (edge-on) near the center, and edge-on (face-on) toward the edges of the disk. Because an edge-on dust grain emits more polarized light and because the light from face-on grains is essentially unpolarized (see Equation~\ref{eq:pprof}), $k$-RAT ($B$-RAT) predicts larger polarization toward the edges (center) of the disk. In the next section, we will perform 3D radiative transfer simulations and use this difference in the polarization profiles to identify the underlying grain-alignment mechanism in \betapic{}.

\section{Synthetic observations}
\label{sec:synobs}
Here we use RADMC-3D to perform a synthetic observation that fits our observation well. We set up a disk model in spherical polar coordinates and assume small dust grains with a single size $a=13.8\rm\, \mu m$,\footnote{It is possible for dust grains to have a distribution of different sizes; however, having a range of different (small) grain sizes does not impact the polarization profile.  Furthermore, the effect of multiple grain sizes on the opacity is compensated for by the universal scaling factor for dust density.} which corresponds to a size parameter $x=0.1$. We assume small dust grains in the dipole regime for their well behaved phase function (i.e., the grains' [polarized] cross section as a function of scattering angle) and ease of computational effort.
Larger dust grains show very complicated phase functions, and the calculation of their optical properties is much more costly; we leave the exploration of models with large dust grains for later work. To compensate for the effects of different dust opacities, we allow the dust density to scale by the same factor across the whole domain in order to achieve the correct intensity. The optical properties for small dust grains in the dipole regime depend both on the composition of dust grains and on the aspect ratio of the oblate spheroids. However, in the end, only the intrinsic polarization fraction $p_0$ matters. We thus choose to fix the composition of dust grains and to vary only the aspect ratio $s$. We adopt the dust model from \cite{Birnstiel2018}, which is a mixture of 20\% water ice \citep{Warren2008}, 33\% astronomical silicates \citep{Draine2003}, 7\% troilite \citep{Henning1996}, and 40\% refractory organics \citep{Henning1996} by mass. Note that this dust model is designed for dust in protoplanetary disks, not for a debris disk like \betapic. However, the choice of dust composition does not affect the polarization profile, which is the main focus of this paper.

We first calculate the dust temperature assuming spherical dust grains (this temperature is also used in Section~\ref{sec:rt}). 
We calculate the optical properties of the dust grains assuming perfect alignment and an oblate-spheroidal geometry with an aspect ratio $s$.
We then calculate the full Stokes parameters assuming oblate grains aligned via the $k$-RAT mechanism, i.e., with the short axes of the dust grains aligned in the radial direction.
We then smooth image with using the $\sim$\,1$\arcsec$ synthesized beam from the ALMA observations.
Since the observed averaged polarization fraction depends roughly linearly on the intrinsic polarization fraction (see Equation~\ref{eq:pprof}), and thus depends monotonically on the aspect ratio $s$, we can easily obtain the best model through a simple binary search in the parameter space of $s$. We do so with a minimum change in $s$ of $0.0001$.

\begin{figure}
    \centering
    \includegraphics[width=0.48\textwidth, clip, trim=0.8cm 0.6cm 0.8cm 0cm]{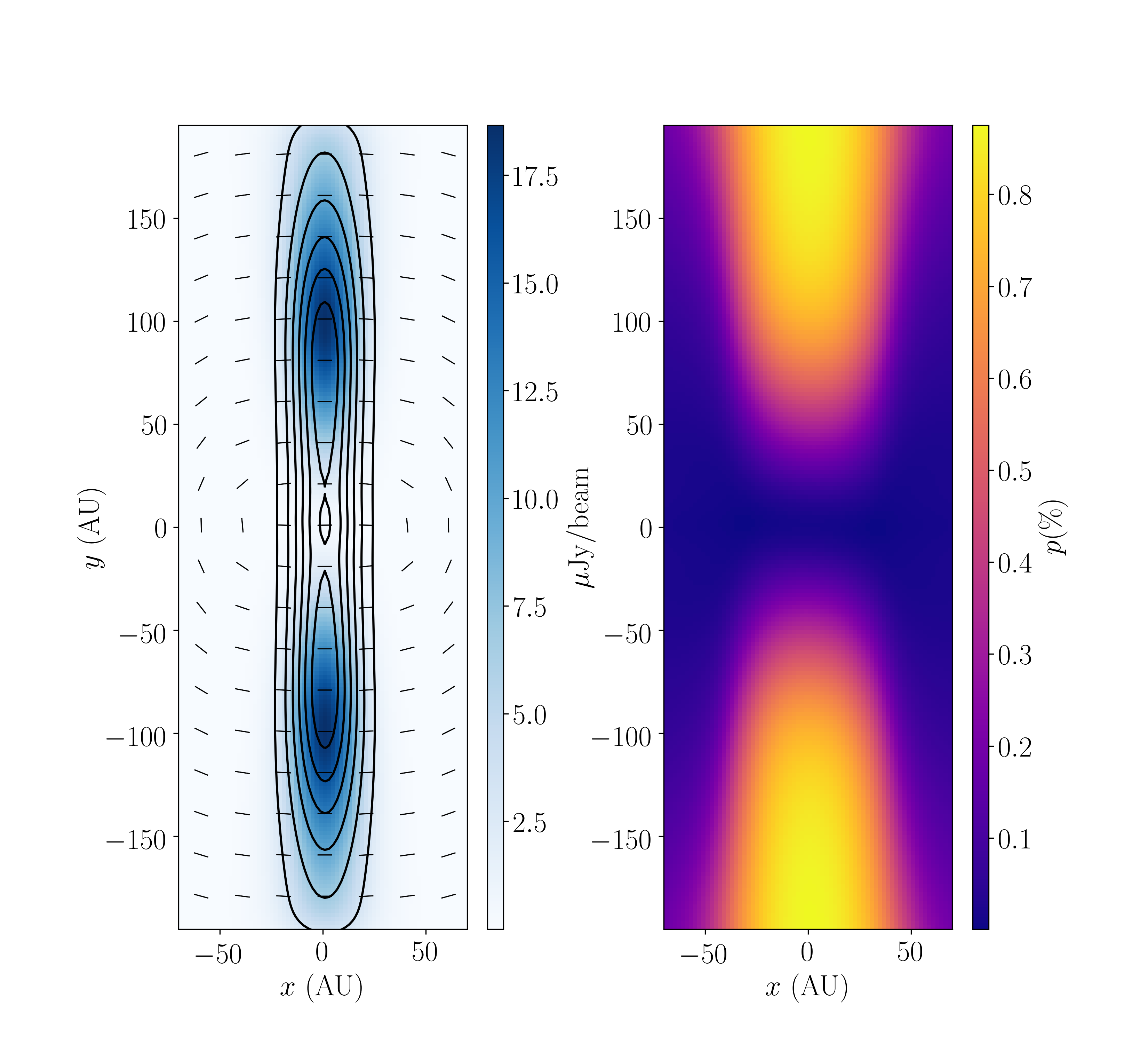}
    \caption{A synthetic observation assuming grains aligned via $k$-RAT. The grain size is $13.8\rm\, \mu m$, which corresponds to a size parameter of $x=0.1$. \textit{Left:} contours are total intensity. Color scale is polarized intensity in \ujybm{}. The line segments are of the same length and represent the polarization orientation. 
    \textit{Right:} the color scale represents the polarization fraction. 
    }
    \label{fig:synobs}
\end{figure}

\begin{figure}
\centering
    \includegraphics[width=0.48\textwidth, clip, trim=0.8cm 0.6cm 0.8cm 0cm]{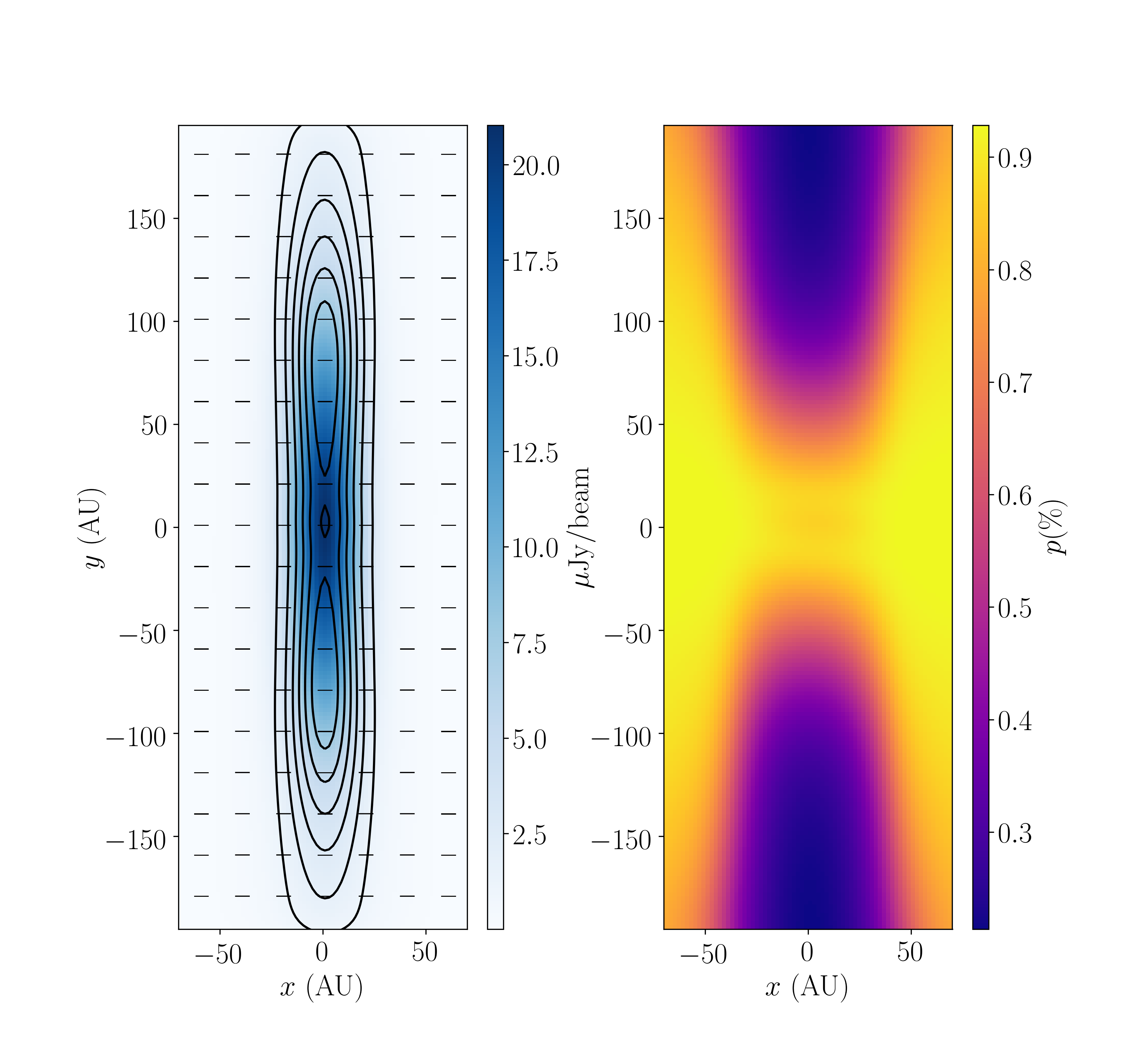}
    \caption{Same as Figure~\ref{fig:synobs} but for $B$-RAT. }
    \label{fig:brat}
\end{figure}

In Figure~\ref{fig:synobs} we show the best-fit synthetic observation assuming $k$-RAT, which has a dust aspect ratio of $s=1.0171$ and an intrinsic polarization fraction $p_0=1.2\%$. Averaging the Stokes parameters from this synthetic observation yields an averaged polarization fraction of $0.50\%$, which matches our full-disk-averaged observations very well. For comparison, we conduct the same calculation assuming $B$-RAT, i.e., where grains are aligned with their short axes along the toroidal direction. In Figure~\ref{fig:brat} we show the best-fit $B$-RAT model, which has $s=1.0125$ and $p_0=0.90\%$, and yields an averaged polarization fraction of $0.51\%$. We can clearly see the difference between these two models: the $k$-RAT model has two off-center peaks of polarization, whereas the polarization in the $B$-RAT model is concentrated near the center of the disk.

\begin{figure}
\centering
    \includegraphics[width=0.47\textwidth]{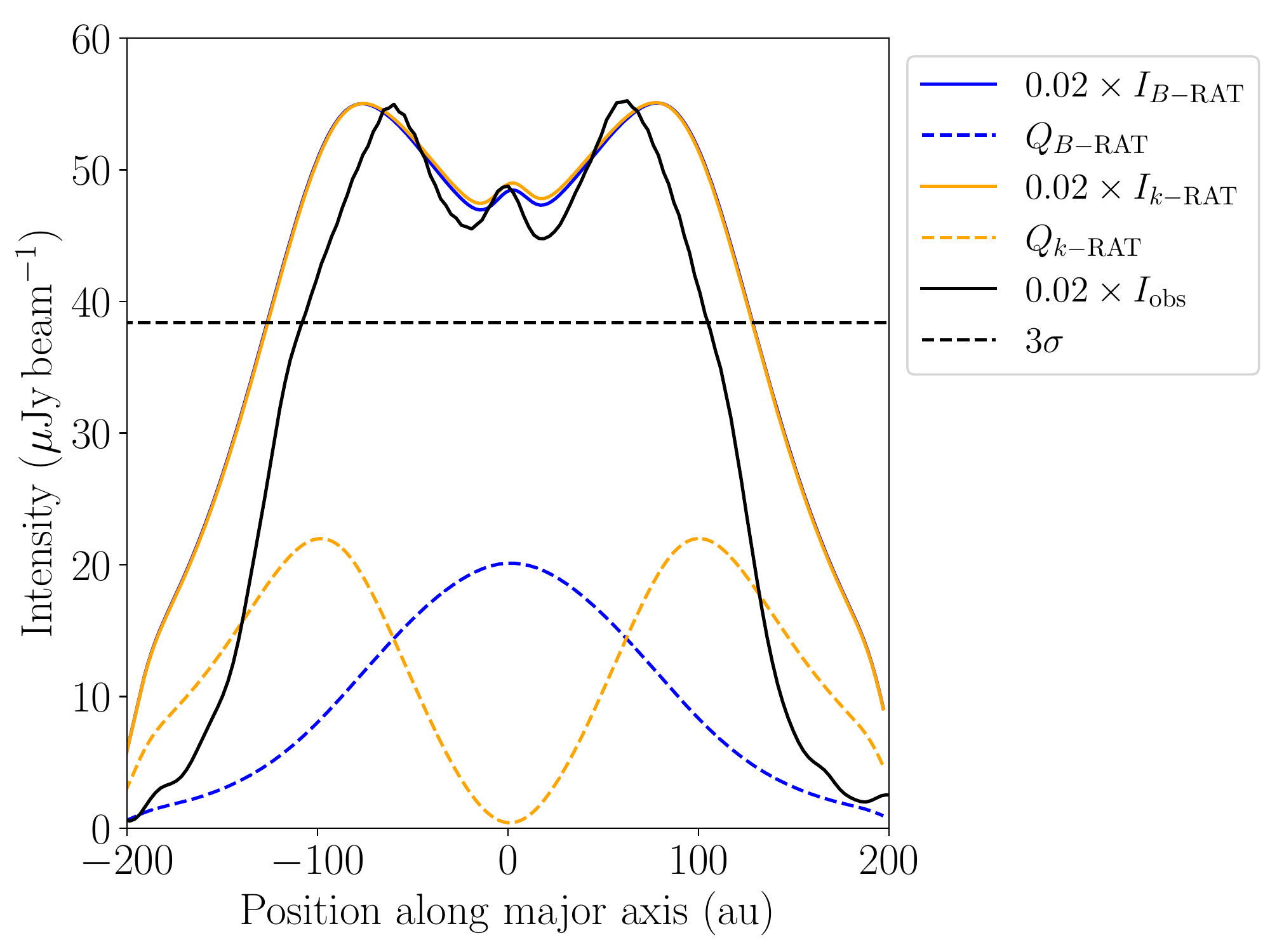}
    \caption{Cuts along the major axis of the modeled and observed disk of \betapic. Blue, orange, and black curves represent 
    the $B$-RAT model, $k$-RAT model, and our observations, respectively. Solid lines represent the
    total intensity, while the dashed lines represent the polarized intensity. To represent the polarized intensity from the ALMA observations, we plot a 3\,$\sigma$ upper limit as a dashed line, where 
    $\sigma$\,=\,\rmsQU\,\ujybm is the rms noise level in the debiased $P$ map. 
    }
    \label{fig:comp}
\end{figure}

To compare these two models in more detail, in Figure~\ref{fig:comp} we plot total intensity and polarized intensity cuts along the disk midplane for both the data and the models. The blue and orange curves represent results from the $B$-RAT and $k$-RAT models, respectively. We use solid lines to show total intensity profiles and dashed lines to represent polarized intensity $Q$ (note that $U=0$ due to the symmetry of the system). We multiply the total intensity by $0.02$ to show both the polarized and non-polarized intensities clearly on the same scale. We also plot our observed total intensity as a black curve, along with a straight black dashed curve showing the 3\,$\sigma$ upper limit in the debiased polarized intensity $P$, where $\sigma$\,=\,\rmsQU\,\ujybm. We can see that both models predict polarized intensity levels below the noise level in the resolved map. They also yield full-disk-averaged polarization fractions similar to our detected value ($0.51\%$). Hence, we cannot distinguish the two models using only the full-disk-averaged value of the polarization fraction from our observations. 

\begin{figure}
    \centering
    \includegraphics[width=0.5\textwidth]{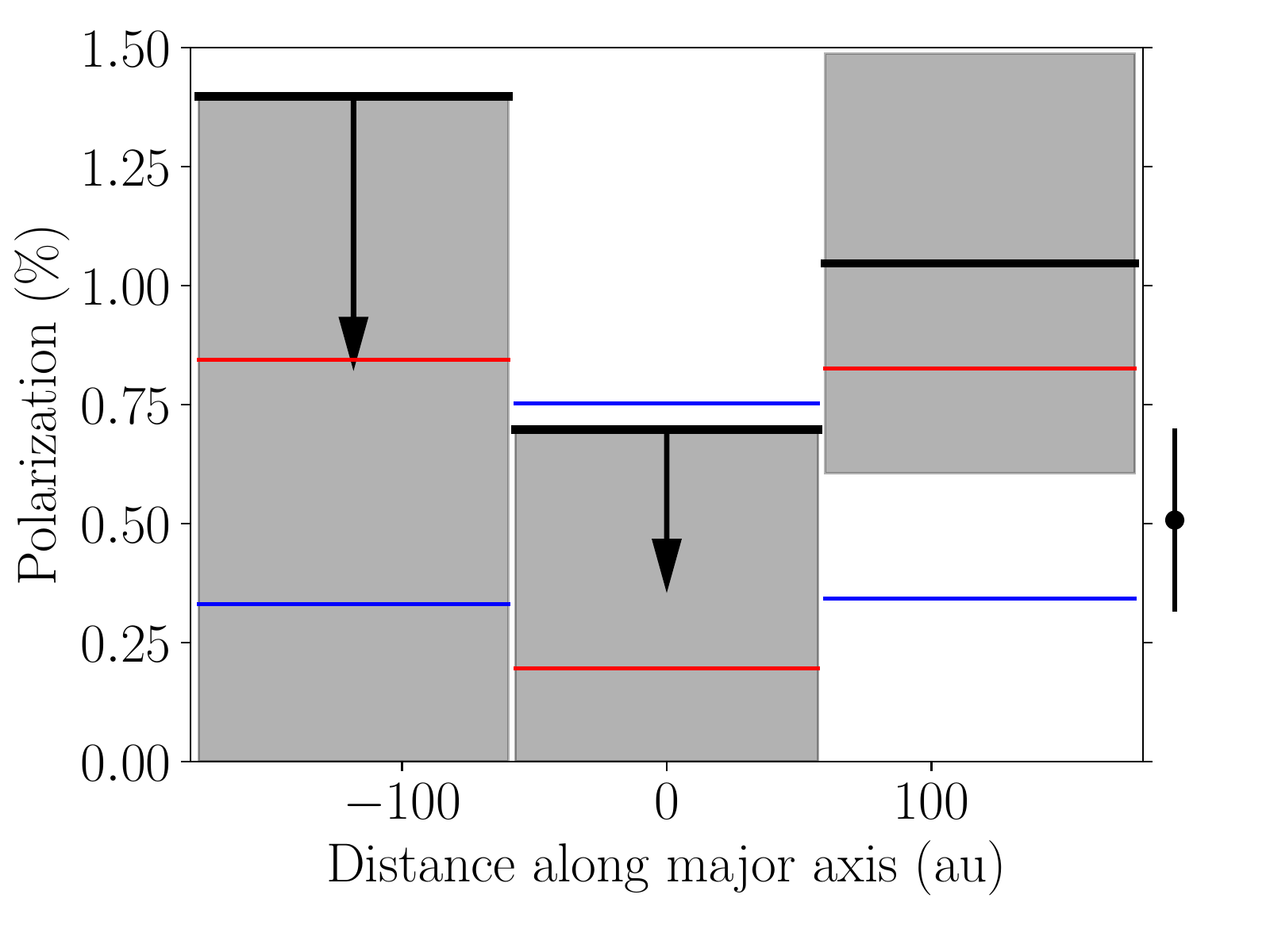}
    \caption{Box-averaged polarization fractions with errors. 
    The $x$-axis ranges from the NE (negative) to the SW (positive); the width of the three bands represents the coverage of each box along the major axis of the disk.
    The red and blue lines show the box-averaged polarization fractions in the $k$-RAT and $B$-RAT models, respectively.
    We treat the observational results, which we plot as thick black lines, in two different ways. For the NE and central thirds, we plot the $3\,\sigma$ upper limits with downward-pointing arrows. For the SW third, where we have a marginal detection of polarized dust emission, we plot our detected polarization fraction, surrounded by a gray box showing $\pm\,1\,\sigma$ error. 
    Just to the right of the plot we show a dot with $\pm\,1\,\sigma$ error bars representing the averaged polarization fraction in the whole disk (\Pfracpctboxavg\%).}
    \label{fig:3box}
\end{figure}

To distinguish between $k$-RAT and $B$-RAT, we use three boxes that covers the NE third, center third, and the SW third of the disk separately (see Section \ref{sec:divided} for a description of the averaging box). All boxes have a width along the minor axis of \boxminorarcsec, or \boxminorau\,au, the same as the full-disk averaging box. The polarization fraction values from our ALMA observations and from the two synthetic observations, measured in each of the three averaging boxes, are shown in Figure~\ref{fig:3box}. We see that $B$-RAT can be excluded because it predicts too much polarization in the center box, larger than the $3\,\sigma$ upper limit in that box. It also predicts polarization that is too weak in the SW box; the value from the model is more than $1\,\sigma$ lower than the marginal detection (SNR\,=\,\SNRPSWthird) of $P$ in the ALMA data reported in Section \ref{sec:divided}. On the other hand, while the $k$-RAT model fails to predict the asymmetry between the NE and SW thirds, the model otherwise agrees very well with the observations: the predicted $k$-RAT polarization in the NE and central boxes lies below $3\,\sigma$ upper limits from the observations, and the polarization in the SW third is within $\pm \,1\,\sigma$ of our tentative detection.  In short, \textit{We find that $k$-RAT is the likely mechanism producing the polarized emission in \betapic.} 

\section{Constraints on dust models}
\label{sec:constraints_dust}

In Section~\ref{sec:rt} we derive constraints on the intrinsic polarization fraction $p_0$ of the dust grains. We find that $p_0$ must be $<1.7\%$ in order to explain our non-detection in the spatially resolved map.
While $p_0$ is directly connected to the observed polarization, which scales approximately linearly with $p_0$, it is not directly connected to the dust models. In this section we discuss constraints on our dust models, regarding both the geometry of the grains and the degree of grain alignment. 
The dependence of $p_0$ on the dust composition is complex and is expected to be secondary to the dependence of $p_0$ on the dust aspect ratio $s$. Consequently, we fix our dust composition and adopt the dust model from \citet{Birnstiel2018}, as before.

\begin{figure}
    \centering
    \includegraphics[width=0.5\textwidth]{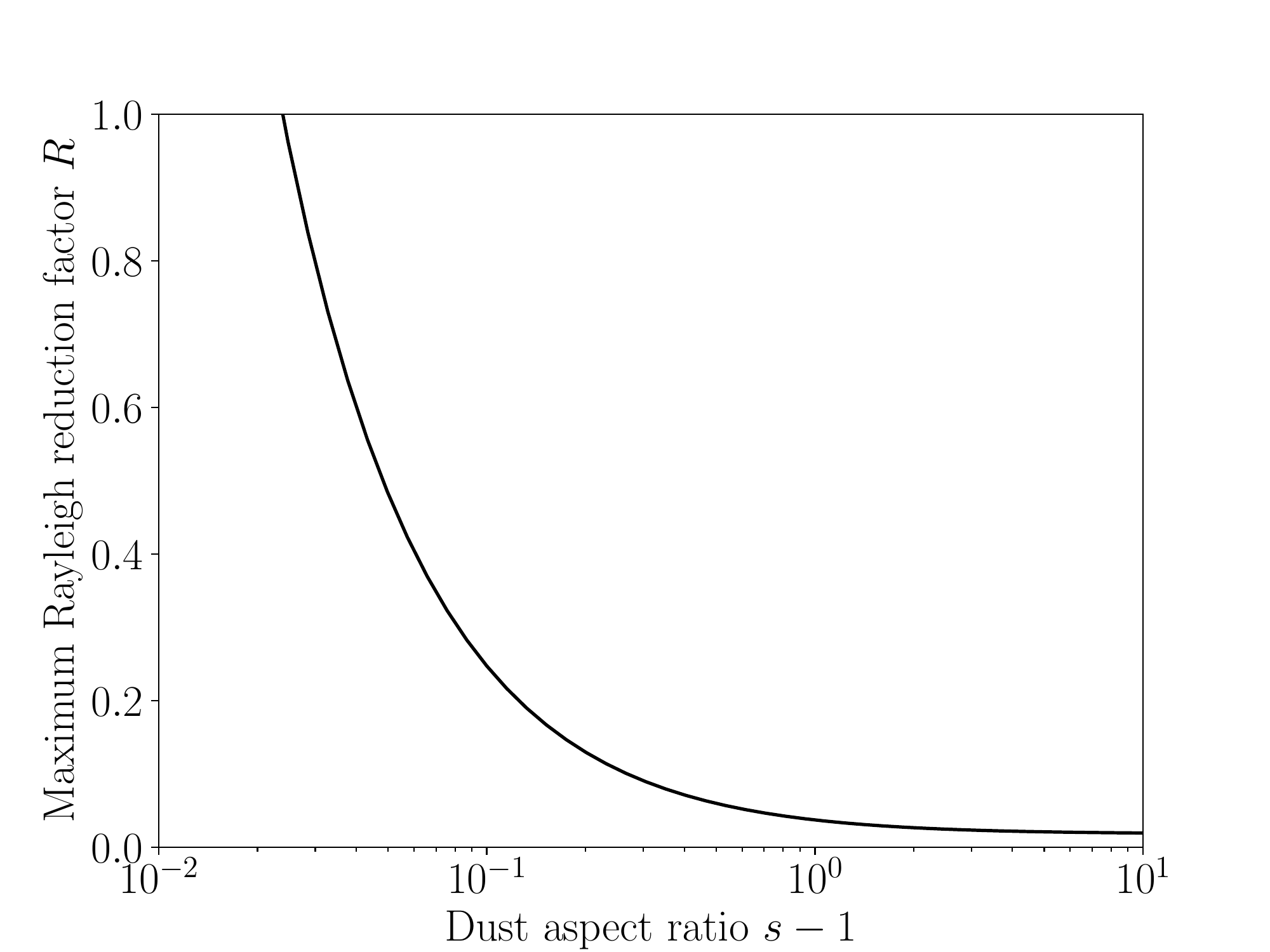}
    \caption{The constraint on the Rayleigh reduction factor ($Rp_0<1.7\%$) as a function of aspect ratio $s$ for small, oblate-spheroidal dust grains. See text for details.}
    \label{fig:ps}
\end{figure}

We first calculate the intrinsic polarization fraction $p_0$ as a function of the aspect ratio $s$ for perfectly aligned, small, oblate-spheroidal dust grains. Since the product of the Rayleigh reduction factor and the intrinsic polarization fraction $Rp_0$ is what determines the observed polarization fraction (see discussion in Section~\ref{sec:rt}), we can divide the upper limit of $1.7\%$ by $p_0$ to get an upper limit for the Rayleigh reduction factor $R$. We plot the results in Figure~\ref{fig:ps}.
We can see that grains with aspect ratio $s<1.025$ can be perfectly aligned with $R=1$ while producing polarization below our detection limit in the resolved map.
However, those aspect ratios are close to unity (i.e., grains that are nearly round), which is unreasonable. For comparison, the disk of Saturn has an aspect ratio of approximately $1.1$ \citep{Saturn} and a sample of asteroids imaged by Gaia have an average aspect ratio of 1.25 \citep{Mommert2018}.  It is hard to imagine dust grains having such small aspect ratios after collisional fragmentation; and indeed, interplanetary dust particles from our Solar system, which should be similar to the dust grains in \betapic, show visibly non-circular aspect ratios \citep{Bradley2003}. 
For more realistic dust grains with larger aspect ratios, perfect alignment models are rejected. In order for their polarized emission to remain undetected, these more elongated dust grains would need to be aligned with low grain-alignment efficiency. The maximum permitted $R$ drops quickly as we increase $s$. For example, when $s=1.1$, we have $R<0.2$, indicating low alignment efficiency. In the limit of $s\gg 1$, the intrinsic polarization fraction $p_0=1$, corresponding to an upper limit of $R<0.017$. We will discuss this result in the context of grain alignment theory in Section~\ref{sec:theory}.  

One caveat of the above discussion is that we assume compact dust grains. \cite{LiGreenberg1998} modeled the \betapic{} disk with a comet dust model and found that the dust grains are highly porous, with a porosity of around $0.95$. The intrinsic polarization of porous dust grains was recently studied by \cite{Kirchschlager2019}, who found that $p_0$ can be reduced by about a factor of $5$ for dust grains with a porosity of $0.8$. Our aforementioned constraints on the degree of alignment $R$ can be loosened by the same factor. For example, if we have dust grains with an aspect ratio $s=1.5$ and a porosity of $0.8$, the degree of alignment $R<0.3$, whereas compact dust grains with $s=1.5$ have $R<0.06$.

Another major caveat is the assumption of small grains. \cite{Cho2007} showed that the intrinsic polarization fraction is very low for large grains. It is also possible that the alignment degree of very large grains ($>1\rm\, mm$) is different from (and very small compared with) the alignment degree of small grains. In both cases, large grains do not contribute polarized emission, but still contribute to the total intensity (Stokes $I$) emission. It is possible that there are well aligned small grains alongside large dust grains in \betapic, and thus the aforementioned constraint on the alignment degree of the small grains would be less stringent (i.e., the degree of alignment could be higher than our estimate shown in Figure~\ref{fig:ps}). We leave detailed explorations of models with large dust grains for future studies.

\section{Dust-grain alignment analysis}
\label{sec:theory}
\subsection{Models for the magnetic field and the radiation field}
\label{ssec:gadisk_model}
The energy density and anisotropy of the radiation field are important for the theory of dust-grain alignment via the RAT mechanism, which is currently the favored mechanism for grain alignment. There are three major contributions to the radiation energy density: thermal emission from dust, the cosmic microwave background (CMB), and stellar illumination. 
\citet{Kral2017} calculated the far-infrared energy density in the radiation field of the \betapic debris disk and found that the energy density of thermal dust emission is greater than that of the CMB. We consequently ignore the CMB in this work.
The thermal dust emission peaks around $50\rm\, \mu m$ and has a total energy density of 
$\nu F_\nu\approx 2\times 10^{-7} \rm\, erg\,cm^{-3}$.
The energy density distribution is relatively flat within the inner $100$\,au of the disk and falls off at larger radii roughly as $r^{-4}$. The anisotropy of the radiation field $\gamma$ from thermal dust emission is typically on the order of $0.1$ \citep{Tazaki2017}; we use this value in our analysis.

\begin{figure}
    \centering
    \includegraphics[width=0.48\textwidth, clip, trim=0.5cm 0cm 0cm 0cm]{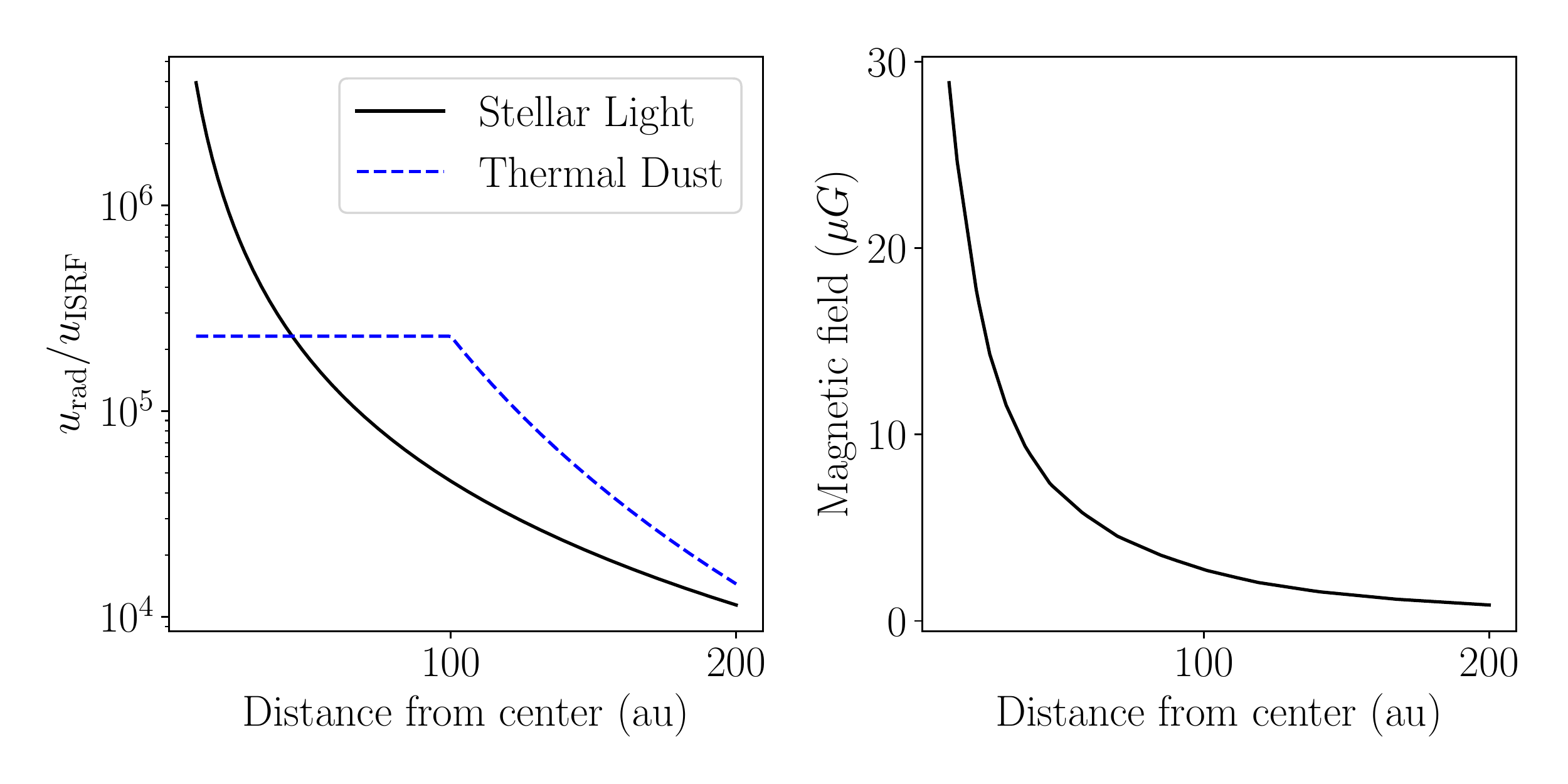}
    \caption{\textit{Left:} The energy density profiles of the stellar radiation and thermal dust emission. \textit{Right:} The magnetic field-strength profile assumed in our grain alignment analysis.}
    \label{fig:gamodel}
\end{figure}

Close to the central star \betapic, the stellar illumination is much stronger than the dust thermal emission.
For a central star with a bolometric luminosity $L = 8.7\rm\, L_\odot$, the radiation energy density is:
\begin{equation}
\begin{split}
u&=\frac{L}{4\pi r^2 c} \\
&= 3.95\times 10^{-4}\, r_\mathrm{au}^{-2} \, \mathrm{erg\,cm^{-3}}\,\,,
\end{split}
\end{equation}
where $r$ is the distance from the central star and $r_\mathrm{au}$ is the distance in au.
The stellar illumination is purely radial. We assume that the anisotropy of the radiation field for this component is $\gamma=1$. The energy-weighted average wavelength, defined as 
$\bar{\lambda}\equiv \int u_\lambda \lambda d\lambda \,\,/ \int u_\lambda d\lambda$,
is $\bar{\lambda}=0.66\rm\, \mu m$ for an effective temperature of $8052$\,K. We plot the energy density profiles for both the stellar radiation and the thermal dust emission in the left panel of Figure~\ref{fig:gamodel}. 

Regarding the magnetic field strength, \citet{KralLatter2016} argue that non-ideal MHD effects are subdominant in the \betapic debris disk and that the MRI is likely to be operating in the disk. It is well known that the MRI is suppressed if the magnetic field is too strong, i.e., when the plasma $\beta \lesssim 10$, where $\beta$ is defined as the ratio of thermal pressure to magnetic pressure \citep[e.g.,][]{Davis2010}. For simplicity, we adopt a magnetic field model assuming $\beta=10$ everywhere. We plot the resulting profile of the magnetic field strength in the right-hand panel of Figure~\ref{fig:gamodel}.  The field strength of our model at a radius of 50\,au is approximately 5\,$\mu$G, which is similar both to the strength of the interplanetary magnetic field measured by Voyager 1 and 2 at a similar distance from our Sun \citep{Burlaga2003} and to the 6\,$\mu$G strength of the magnetic field in the cold neutral interstellar medium of the Milky Way \citep{Heiles2005}.

\subsection{Alignment timescales}
\label{ssec:timescale}

The alignment of dust grains is a complicated process. We first compare several timescales involved in the grain-alignment process following the work presented in \citet{Tazaki2017} and \citet{Yang2021}. Three of the most important timescales are the gas damping timescale, the Larmor precession timescale, and the RAT precession timescale.

The gas damping timescale is the timescale on which random bombardment by gas particles misaligns the dust grains. In our disk model the gas component is assumed to be a mixture of carbon and oxygen atoms (or their ions; electrons are ignored in the damping process). The mean molecular weight is $14$, as opposed to the usual $2.34$ for protoplanetary disks dominated by molecular hydrogen. The gas damping timescale $t_g$ is as follows:
\begin{equation}
    \begin{split}
    t_g = 2.3\times 10^{6}~\mathrm{yr}
    &\times \hat{\rho}_s
    \left(\frac{a}{5~\mathrm{\mu m}}\right)\\
    &\left(\frac{n_{\rm g}}{20~\mathrm{cm^{-3}}}\right)^{-1}\left(\frac{T_g}{85~\mathrm{K}}\right)^{-1/2},
    \end{split}
    \label{eqn:gas_damping}
\end{equation}
where $\hat{\rho}_s=\rho_s/(3{\rm\, g\,cm^{-3}})$ is the reduced solid density of the dust grain, which we take as $\hat{\rho}_s=1$; $a$ is the effective radius of the dust grain; $n_g$ is the number density of gas particles; and $T_g$ is the gas temperature.

As a result of the Barnett effect, a rotating dust grain has a magnetic moment proportional to its angular momentum \citep{Barnett1915}. Hence, there is a Larmor precession timescale $t_L$ defined as the precession period of dust grains around an external magnetic field:
\begin{equation}
    \begin{split}
    t_L = 2.1\times 10^2\,\mathrm{yr} 
    &\times \hat{\chi}^{-1} \hat{\rho}_s
    \left( \frac{T_d}{15~\mathrm{K}} \right) \\
    &\left( \frac{B}{5~\mathrm{\mu G}} \right)^{-1} \left( \frac{a}{5~\mathrm{\mu m}} \right)^2,
    \end{split}
    \label{eq:tL}
\end{equation}
where $T_d$ is the dust temperature, $B$ is the magnetic field strength, and $\chi=10^{-3}\,\hat{\chi}\,(T_d/15\mathrm{\, K})^{-1}$ is the magnetic susceptibility of the dust grain; for regular paramagnetic materials, $\hat{\chi}\approx 1$. If dust grains contain clusters of ferromagnetic material, known as ``superparamagnetic inclusions,'' $\chi$ can be enhanced by up to a factor of $10^3$ \citep{JS1967,Yang2021}. 

The third timescale is the RAT precession timescale. Radiation can exert torques (i.e., RATs) on dust grains that have a significant helicity. The RAT precession timescale $t_{\rm rad,\,p}$ is the precession period of the dust grains experiencing such torques:
\begin{equation}
\begin{split}
t_{\rm rad,\,p} &= 4 \times 10^{-3}\,\mathrm{yr}\times \hat{\rho}_s^{1/2} \hat{s}^{-1/3} \left(\frac{a}{5\rm\, \mu m}\right)^{1/2} \left(\frac{T_d}{15\rm\, K}\right)^{1/2}\\
&\left(\frac{u_\mathrm{rad}}{2\times10^5 u_\mathrm{ISRF}}\right)^{-1} \left(\frac{\bar{\lambda}}{0.66\rm\, \mu m}\right)^{-1}
\left(\frac{\gamma \overline{|Q_\Gamma|}}{0.04}\right)^{-1},
\end{split}
\label{eq:trad}
\end{equation}
where the radiation field energy density $u_\mathrm{rad}$ is normalized by $u_\mathrm{ISRF}=8.64\times10^{-13}\rm\, erg\,cm^{-3}$, the energy density of the standard interstellar radiation field  \citep[ISRF;][]{Mathis1983,LazarianHoang2007a}. 
$\overline{|Q_\Gamma|}$ is a dimensionless parameter describing the strength of the torque that follows the following piecewise function \citep{LazarianHoang2007a}:
\begin{equation}
|Q_\Gamma| \approx \left\{
\begin{array}{ll}
2.3\left(\frac{\lambda}{a}\right)^{-3} &\quad \mathrm{for}\ \lambda>1.8\,a \\
0.4 &\quad \mathrm{for}\ \lambda\leq 1.8\,a
\end{array}
\right..
\end{equation}
Since we have both stellar illumination and thermal dust emission, we calculate the RAT precession timescale for these two components separately according to Equation~\eqref{eq:trad}, and then take the harmonic average of these two results to obtain the RAT precession timescale when both sources are in action.

Although other grain-alignment mechanisms exist, we focus only on the RAT alignment theory. Grain alignment via the Gold mechanism \citep{Gold1952} and Mechanical Alignment Torques \citep[MATs;][]{LazarianHoang2007b} both rely on differential motion between the gas and the dust to operate; both of these mechanisms should be less effective than RATs at aligning grains in the \betapic debris disk because of the low density of gas particles and correspondingly rare interactions between gas and dust, as shown in Equation~\eqref{eqn:gas_damping}. 

In debris disks, the timescale of collisions between dust grains is  important because it determines the lifetime of dust grains (i.e., how long they survive before being collisionally destroyed). 
If we assume the dust grains are primarily destroyed by collisions with similar sized grains, and we further assume that the dust grains have a power law size distribution $dn/da\sim a^{-3.5}$ with a maximum grain size of $a_2=1$ cm \citep{Zagorovsky2010}, we can derive the collision timescale using Equation (11) of \cite{Ahmic2009} as follows: 
\begin{equation}
t_\mathrm{col} = \frac{5P\rho_s \sqrt{aa_2}}{54\Sigma}\,\,,
\end{equation}
where $P$ is the local Keplerian rotation period and $\Sigma$ is the dust column density (viewed face-on). 

\begin{figure*}[hbt!]
\centering
\includegraphics[width=0.49\textwidth]{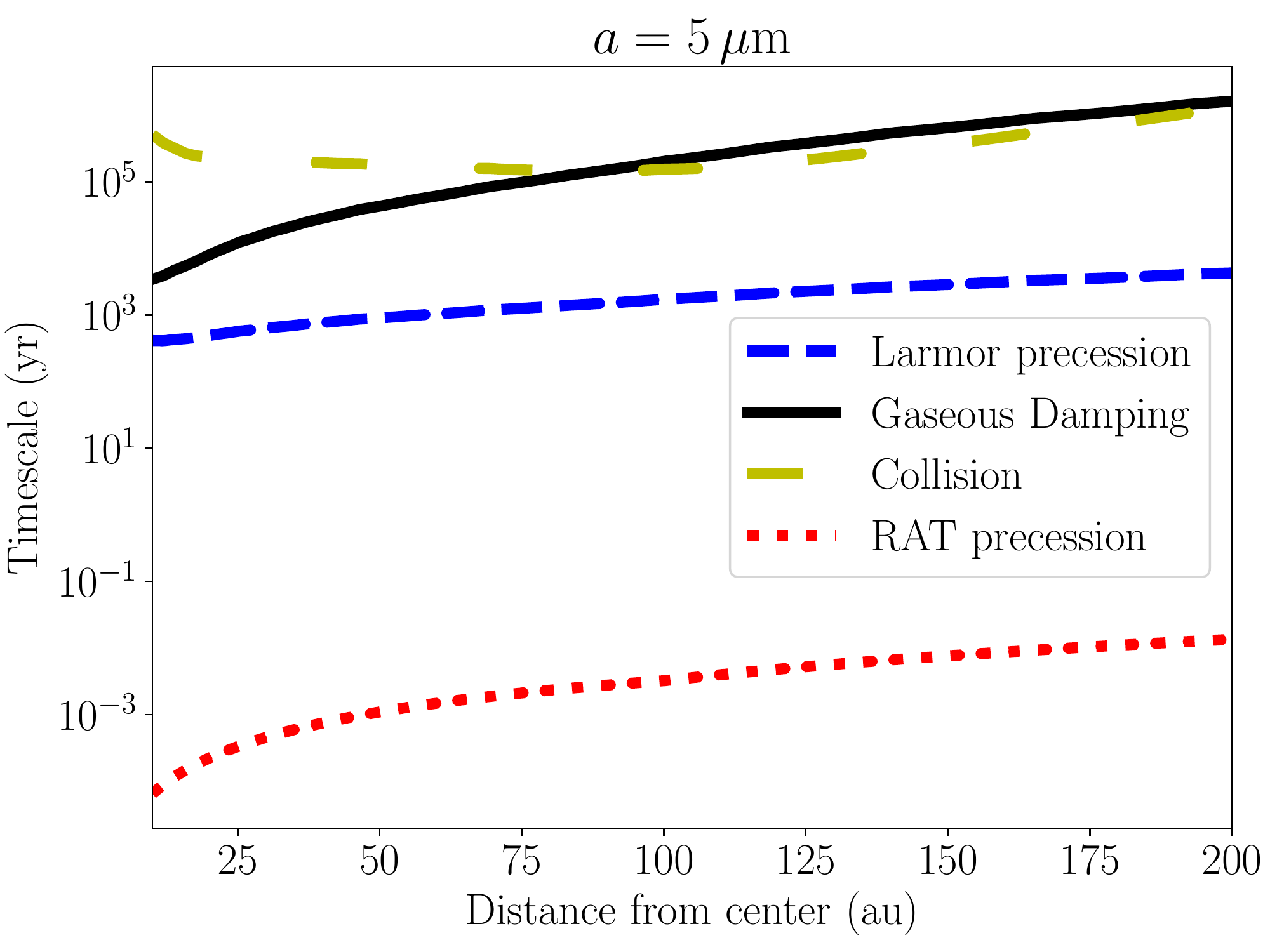}
\includegraphics[width=0.49\textwidth]{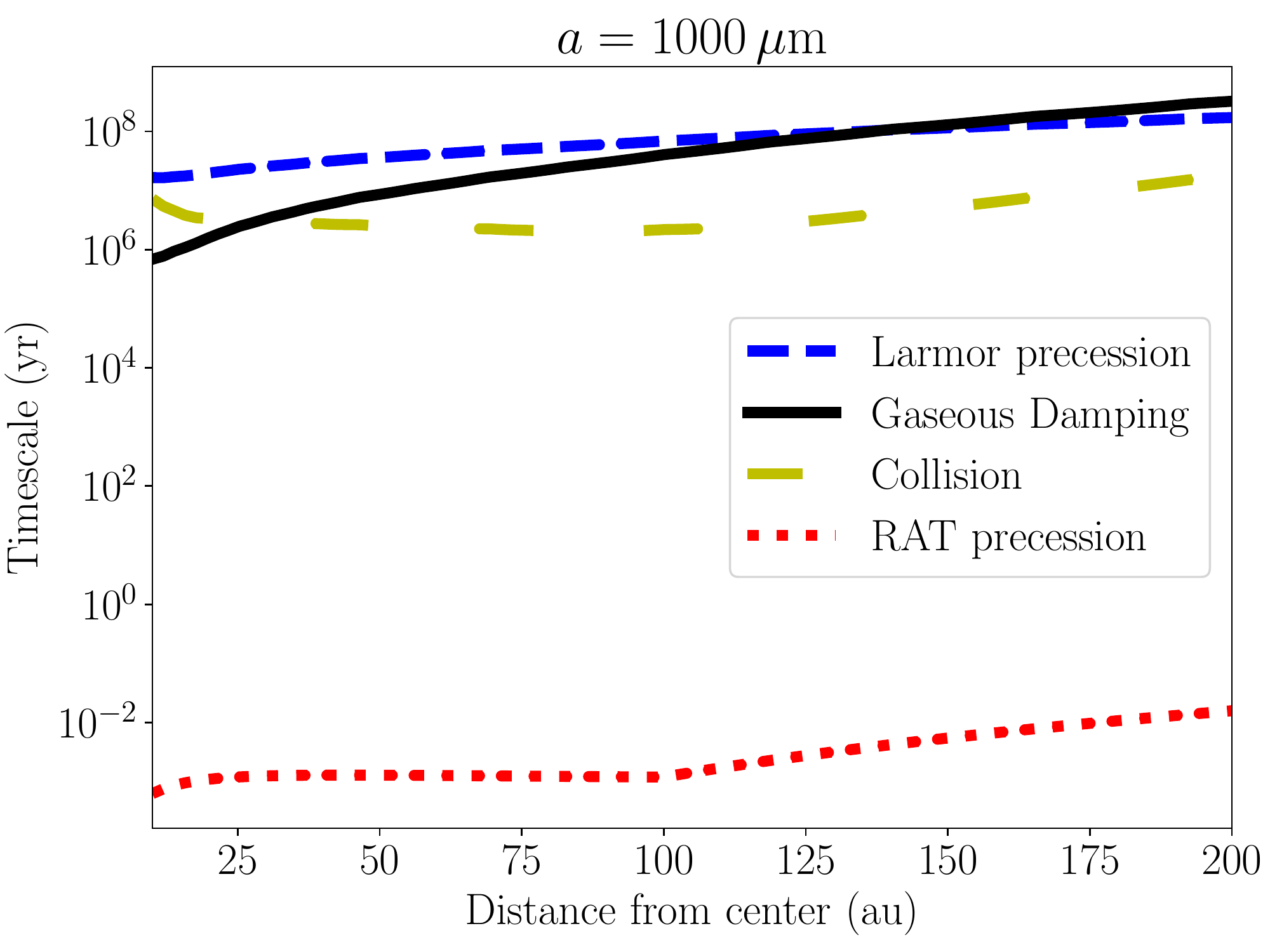}
\caption{Comparison of dust-grain alignment timescales for small dust grains ($a=5\rm\, \mu m$, left panel) and large dust grains ($a=1$\,mm, right panel).}
\label{fig:timescales}
\end{figure*}

In Figure~\ref{fig:timescales} we show a comparison of the four timescales $t_g$, $t_L$, $t_{\mathrm{rad,\,p}}$, and $t_\textrm{col}$ as a function of distance from the center of the \betapic debris disk. In the left panel we show the results for small dust grains with $a=5\rm\, \mu m$, which is close to the blow-out size of dust grains in \betapic \citep{Burns1979, Arnold2019}.  We see that the Larmor precession timescale is always smaller than the gas damping timescale for small grains. In addition, we find that RAT precession is at least 4 orders smaller than the other two timescales, and thus RAT is very likely to be operating. In the right panel, we show the results for large dust grains with $a=1\rm\,mm$. We see that the Larmor precession and gas damping timescales are comparable to one another, especially at the location of the dusty ring at $\sim\,100$\,au. Note that this applies only to regular paramagnetic grains. If we consider superparamagnetic inclusions, the Larmor precession timescale can easily be reduced by a factor of $\sim$\,$10^{2-3}$, which would result in a Larmor precession timescale that is always smaller than the gas damping timescale. In summary, while $t_g$ and $t_L$ can be comparable depending on the grain size we consider, the RAT precession timescale is $4-10$ orders of magnitude smaller than both of them.  There is thus no doubt that the RAT alignment mechanism is operating in this system. 
We also find that the collision timescale $t_\mathrm{col}$ is smaller than both $t_L$ and $t_g$ for $1$\,mm dust grains, which means such dust grains would be destroyed through collisions before they can be aligned with magnetic fields. 

When RAT is operating, there are two possible outcomes. In the presence of a strong external magnetic field, dust grains will become aligned with the magnetic field (the $B$-RAT case). In the absence of a strong external magnetic field, dust grains will still be aligned, but now with respect to the radiation flux instead of with respect to the magnetic field (the $k$-RAT case).  The strength of the magnetic field is quantified by the Larmor precession timescale $t_L$. As suggested by \cite{LazarianHoang2007a} and \cite{Tazaki2017}, in the regime where $t_\textrm{rad,p}\ll t_L$, we should have $k$-RAT alignment instead of $B$-RAT alignment. 
And indeed, in order for $B$-RAT to dominate over $k$-RAT in \betapic at a distance of, e.g., $100$\,au, the strength of the magnetic field would need to be $\sim 1.5\rm\, mG$ for $5\rm\, \mu m$-sized dust grains and $\sim 160\rm\, G$ for $1$\,mm grains, in both cases considering grains with superparamagnetic inclusions with $\hat{\chi}=10^3$ \citep{Yang2021}. Given the microgauss-level magnetic field strengths discussed in Section \ref{ssec:gadisk_model}, these high values for the magnetic field in \betapic are unrealistic. In summary, $k$-RAT is the likely mechanism producing the polarization we see in \betapic. This is in agreement with the analysis of both our ALMA observations and our synthetic observations.

\subsection{Degree of Alignment}

Even though the above timescale comparisons show that RAT is operating very well, it does not guarantee perfect alignment of dust grains with respect to the magnetic or radiation field. There are several other factors that come into play to determine the degree of alignment, which equals the Rayleigh reduction factor $R$.

The main one is the internal alignment, i.e., the alignment of the dust grain's angular momentum $J$ with the grain's principle axis of maximum inertia. \cite{Tazaki2017} considered several relevant relaxation processes including Barnett relaxation of both electrons and nuclei, Barnett relaxation with superparamagnetic inclusions, and inelastic dissipation. The results are shown in their Figure 2. For grains larger than $5\rm\, \mu m$, the internal relaxation timescale, $t_\mathrm{int}$, is longer than $\sim 3\times 10^6\rm\, yr$. The internal relaxation timescale increases rapidly as grain size $a$ increases (increasing roughly as $a^7$). We can thus conclude that the internal relaxation timescale is much longer than the other relevant timescales in the entire \betapic debris disk, and thus grains are not internally aligned. 

Without internal alignment, grain alignment is still possible, as shown by \cite{Hoang2009}. This is especially true when there are so-called ``high-$J$ attractors,'' which are attracting stationary points in the phase diagram (whose axes are the angle with the aligning field $\theta$ and the angular momentum $J$; see, e.g., \citealt{LazarianHoang2007a}) where grains have suprathermally rotating angular momentum. Grains at high-$J$ attractors are expected to be aligned perfectly, regardless of internal relaxation \citep{Hoang2009}. Grains at low-$J$ attractors are subject to thermal fluctuations, and can also be aligned in the ``wrong'' configuration, i.e., with their short axes not along the direction of the aligning field; in this case the degree of alignment is strongly affected by the lack of internal relaxation \citep{Hoang2009}. Whether high-$J$ attractors exist depends on the geometry, composition, and size of the dust grains. 

Let $f_{\mathrm{high-}J}$ be the fraction of dust grains aligned at high-$J$ attractor points. The timescale comparison in Section~\ref{ssec:timescale} shows that we are in a regime where $t_\mathrm{rad,p}<t_g<t_\mathrm{int}$. For grains aligned at high-$J$ attractors, we have perfect alignment. For grains not aligned at high-$J$ attractors, the degree of alignment is $0$ due to the lack of internal alignment (\citealt{Hoang2009}; see also discussions by \citealt{Tazaki2017}). As a result, $R=f_{\mathrm{high-}J}$. From our previous constraint on the Rayleigh reduction factor (see Section \ref{sec:constraints_dust} and Figure \ref{fig:ps}) we conclude that the $f_{\mathrm{high-}J}$ is likely to be very small. For dust grains with aspect ratios $s > 1.1$, we have $f_{\mathrm{high-}J}<0.2$. For grains with $s>2$, we have $f_{\mathrm{high-}J}<0.037$. These values of $f_{\mathrm{high-}J}$ are far smaller than the values typically adopted in the literature. Recently, \cite{Herranen2021} studied the alignment efficiency of ensembles of Gaussian random ellipsoidal dust grains and found that the alignment efficiency of small grains exposed to the ISRF can reach levels as high as $\sim0.5$.  Additionally, \citet{LeGouellec2020} found evidence for high grain-alignment efficiency in Class 0 protostellar cores.  Whether our inference of low grain-alignment efficiency in \betapic is possible with realistic distributions of dust grains is a question for RAT theory to answer.

\section{Conclusions}
\label{sec:conc}

We present 870\,$\micron$ ALMA polarization observations of thermal dust emission toward the edge-on \betapic debris disk. Our analysis of the observations allows us to draw the following conclusions:

\begin{enumerate}
    \item The spatially resolved maps do not exhibit any detectable dust polarization.
    
    \item When we average the emission across the entire disk in a box of size \boxminorau\,au\,$\times$\,\boxmajorau\,au, we detect polarized dust emission at a marginally significant SNR level of \SNRPboxavg, finding a polarization fraction  $P_\textrm{frac}$\,=\,\Pfracboxavg (i.e., \Pfracpctboxavg\%). The polarization position angle $\chi$\,=\,\PAboxavg$\degree$, i.e., along the minor axis of the disk.
    
    \item To improve the SNR of the observations, we transform the Stokes ($Q$,\,$U$) parameters into the ($Q^\prime$,\,$U^\prime$) frame such that $\pm\,Q^\prime$ correspond to polarization with an orientation along the minor axis ($+$) and along the major axis ($-$) of the disk.  After doing so, we detect $+\,Q^\prime$ at a statistically significant SNR level of \SNRQprimeboxavg; $U^\prime$ is consistent with noise.  
    
    \item When we average the polarized emission across different regions of the disk, we find that the polarization primarily arises from the SW third of the disk; polarized emission is not detected toward the NE or central thirds.  
\end{enumerate}

We compare our observations with models of dust scattering and models of dust grains aligned via the radiative torque (RAT) mechanism both with respect to the radiation flux ($k$-RAT) and with respect to the magnetic field ($B$-RAT).  We find the following:

\begin{enumerate}
\setcounter{enumi}{4}
    \item Polarization from scattering by dust grains can be ruled out due to the low optical depth of the \betapic debris disk at submillimeter wavelengths.
    
    \item With a simple radiative transfer model we constrain the intrinsic polarization fraction of the dust grains to be $p_0<1.7\%$ and $p_0<1.3\%$, assuming small grains aligned via $k$-RAT and $B$-RAT, respectively. Grains with larger $p_0$ would produce significant polarized emission in the resolved map, which we do not detect. 
    
    \item We present synthetic observations in Figures~\ref{fig:synobs} and \ref{fig:brat}, assuming $k$-RAT and $B$-RAT, respectively. Both models match our observed, full-disk-averaged polarization fraction very well.  
    To distinguish between $k$-RAT and $B$-RAT, we average the polarized emission across the three thirds of the disk, as we do with the ALMA observations. $B$-RAT is excluded because it predicts too much polarization in the center third and too little polarization in the SW third (see Figure~\ref{fig:3box}).  We find that $k$-RAT is the likely mechanism producing the polarized emission in \betapic. 
    
    \item Given the constraint on the intrinsic polarization fraction $p_0<1.7\%$, we attempt to constrain our dust models. 
    If the dust grains are small and perfectly aligned, they must have a very small aspect ratio of $s<1.025$.
    For dust grains with realistic aspect ratios ($s>1.1$), the degree of alignment must be smaller than $\sim$\,0.2, implying inefficient alignment of dust grains. 
    
    \item Based on grain-alignment timescale comparisons, the RAT alignment mechanism should be operating very well in \betapic, and $k$-RAT is favored over $B$-RAT. This is consistent with the conclusions we draw from the analysis of the ALMA observations and the synthetic observations.
\end{enumerate}

As the first observational and theoretical analysis of deep submillimeter dust polarization observations toward a debris disk, this work paves the way for many future polarization studies of both \betapic and other debris disks.  One path forward would be to extend our models to the regime of large ($\sim$\,1\,mm) dust grains by analyzing multi-wavelength observations.  Furthermore, given the fact that we were able to detect polarized dust emission when averaging across the entire disk of \betapic, it is likely that we would be able to make a spatially resolved polarization map if the sensitivity of the observations were improved by a factor of $\sim$\,2.  This is achievable in \betapic with a substantial additional investment of Band 7 observation time ($\sim$\,9\,hr on-source).  However, this type of deep continuum polarization observation will become more easily achievable in \betapic and in other (both brighter and fainter) debris disks after the ALMA 2030 wideband sensitivity upgrade is complete \citep{Carpenter2019}.



\section*{acknowledgements}

The authors thank the anonymous referee for the insightful comments.
The authors acknowledge the excellent support of the EA ARC, in particular from Kazuya Saigo, Misato Fukagawa, and \'Alvaro Gonz\'alez.  The authors would particularly like to acknowledge the efforts of EA ARC staff member Toshinobu Takagi, who reduced the polarization data in order to confirm the results of the authors' manual reduction of the data.
The authors thank Ryo Tazaki and Alex Lazarian for fruitful discussions regarding RATs.
C.L.H.H. acknowledges Patricio Sanhueza for the vibrant and helpful discussions.
C.L.H.H. acknowledges the support of the NAOJ Fellowship and JSPS KAKENHI grants 18K13586 and 20K14527.
A.M.H. is supported by a Cottrell Scholar Award from the Research Corporation for Science Advancement.
V.J.M.L.G. acknowledges the support of the ESO Studentship Program.
Z.-Y.L. is supported in part by NSF AST-1815784 and NASA 80NSSC18K1095.
This paper makes use of the following ALMA data: 
ADS/JAO.ALMA\#2019.1.00041.S.  
ALMA is a partnership of ESO (representing its member states), NSF (USA) and NINS (Japan), together with NRC (Canada), MOST and ASIAA (Taiwan), and KASI (Republic of Korea), in cooperation with the Republic of Chile. The Joint ALMA Observatory is operated by ESO, AUI/NRAO and NAOJ.
The National Radio Astronomy Observatory is a facility of the National Science Foundation operated under cooperative agreement by Associated Universities, Inc. 
Some archival images of \betapic were retrieved from the Japanese Virtual Observatory (JVO) portal (\url{http://jvo.nao.ac.jp/portal}) operated by ADC/NAOJ.

\textit{Facilities:} ALMA.

\textit{Software:}   
CASA \citep{McMullin2007}.  
Astropy \citep{Astropy2018}. 
RADMC-3D \citep{Dullemond2012}.
This research made use of APLpy, an open-source plotting package for Python \citep{Robitaille2012}.

\bibliography{ms}
\bibliographystyle{aasjournal}

\end{CJK*}
\end{document}